\documentclass[twocolumn,twocolappendix]{aastex631}

\usepackage{amsmath}
\usepackage{enumitem}
\usepackage{multirow}

\begin{document}

\title{Deciphering the morphological origins of X-shaped radio galaxies: \\
Numerical modeling of Back-flow vs. Jet-reorientation}

\correspondingauthor{GG, BV, CF}
\email{gourab@iiti.ac.in, bvaidya@iiti.ac.in, fendt@mpia.de}

\author[0000-0001-9133-1005]{Gourab Giri}
\affiliation{Department of Astronomy, Astrophysics and Space Engineering, Indian Institute of Technology Indore, 453552,
India}

\author[0000-0001-5424-0059]{Bhargav Vaidya}
\affiliation{Department of Astronomy, Astrophysics and Space Engineering, Indian Institute of Technology Indore, 453552,
India}

\author[0000-0002-3528-7625]{Christian Fendt}
\affiliation{Max Planck Institute for Astronomy, K\"onigstuhl 17, D-69117 Heidelberg, Germany}

%\collaboration{20}{(AAS Journals Data Editors)}



\begin{abstract}  
X-shaped Radio Galaxies (XRGs) develop when certain extra-galactic jets deviate from their propagation path. 
An asymmetric ambient medium (Back-flow model) or complex Active Galactic Nuclei activity (Jet-reorientation model) enforcing the jet direction to deviate may cause such structures. 
In this context, the present investigation focuses on the modeling of XRGs by performing 3D relativistic magneto-hydrodynamic
simulations. We implement different jet propagation models applying an 
initially identical jet-ambient medium configuration to understand distinctive features.
This study, the first of its kind, demonstrates that all adopted models produce
XRGs with notable properties, thereby challenging the notion of a universal model. 
Jet reorientation naturally explains several contentious properties of XRGs, including wing alignment along
the ambient medium's primary axis, development of collimated lobes,
and the formation of noticeably longer wings than active lobes. 
Such XRGs disrupt the cluster medium by generating isotropic shocks and channeling more energy 
than the Back-flow scenario. 
Our synthetic thermal X-ray maps of the cluster medium reveal four `clear' elongated cavities associated with
the wing-lobe alignment, regardless of projection effects, though affecting their age estimation.  
We show depth and geometric alignment of the evolved cavities may qualify as promising characteristics of XRGs, which may be used to disentangle different formation scenarios.
\end{abstract}



\section{Introduction} \label{sec:introduction}
A small fraction of Active Galactic Nuclei (AGN) exhibit the presence of jets (magnetised high-speed plasma outflows) that are ejected
by a central supermassive black-hole (SMBH) to sub-kiloparsec (kpc) or even a few mega-parsec (Mpc) distance, typically propagating along a 
bidirectional straight path \citep{Kellermann1989,Bassani2016,Kharb2019,Dabhade2020}. 
Significant deviation from this predetermined propagation path, on the other hand, has been observed in a small 
population of these extended radio galaxies, resulting in unusual radio morphologies \citep{Leahy1992,Cheung2007,Krause2019,Pandge2022}.

Newer discoveries with telescopes such as LOFAR and MeerKat have uncovered morphological features with even more extreme anomalies, such that they may confound 
the present understanding of the phases of jet evolution \citep{Hardcastle2019,Ramatsoku2020,Rudnick2021,Rudnick2022,Gopal-Krishna2022}. 
Despite the complexity of this zoo of peculiar radio galaxies, they can be divided into two broad categories: 
mirror symmetric sources and inversion symmetric sources \citep{Ekers1982}. 

A widespread consensus exists about the formation process of mirror symmetric sources with tailed 
morphologies that can be understood as resulting from a translational motion of the jet hosting galaxy with respect to the circumjacent 
medium, typically in a galaxy cluster environment \citep{Smolcic2007,Oneill2019,Muller2021}. 
However, the formation process of inversion symmetric sources is still under debate. 
The potential confusion arises from question whether the responsible mechanism results from the asymmetric ambient medium or  from 
the complex AGN activities \citep{Rottmann2002}. 

This latter class of sources exhibit morphologies such as X-, S-, Z-, or even W-like jetted structures \citep{Gopal-Krishna2003,Kraft2005,Hodges-Kluck2012,Lal2019}, 
which have been observed not only in extragalactic sources but also in several micro-quasars as well \citep{Roberts2008,Marti2017}. 
Although the geometry of these structures is difficult to disentangle in low resolution images, their morphologies 
appear now to be distinguishable applying modern high resolution and highly sensitive observations 
\citep{Cotton2020,Bruni2021,Mahatma2023}. 
In this work, we will focus on understanding the formation and evolution process of X-shaped radio galaxies by investigating both 
their peculiar jetted structure as well as the properties of their environment.

X-shaped radio galaxies (XRGs) can be characterised by the presence of two pairs of fairly well collimated radio lobes that are aligned at an angle to each other. The average alignment angle between the two pairs of lobes are typically high ($\sim 75^{\circ}$), giving them a distinct X-shape \citep{Capetti2002,Joshi2019,Bruno2019,Ignesti2020,Mahatma2023}. One pair of jet lobes, where hotspots are often observed (edge-brightened) is known as the active lobe, while the other pair (faint, diffuse, and edge-darkened) was formerly believed to have evolved passively, is known as the wing \citep{Hodges-Kluck2011,Giri2022A}.

Due to their peculiar geometry and emission characteristics, several formation mechanisms have been proposed. 
These can be broadly categorised into two basic classes in which the underlying mechanism generating such ambiguities is attributed to: 
(a) the triaxiality of the ambient medium shaping the backflowing material by deflecting them in the lateral direction (Back-flow model; \citet{Leahy1984,Capetti2002,Hodges-Kluck2011,Rossi2017}); or
(b) a complex AGN activity causing the jet-ejection axis to shift or flip from the initial configuration (Jet re-orientation model; \citet{Begelman1980, Dennett-Thorpe2002, Merritt2002,Babul2013,Horton2020}).

There are evidences that support each of these unique models, in addition to significant caveats that have also been discovered. 
The Back-flow model, for example, naturally explains the statistically observed connection of wings to be oriented along the minor axis of the tri-axial ambient medium \citep{Capetti2002,Hodges-Kluck2010A,Gillone2016}; however, a small number of such galaxies, recently discovered, have been observed to deviate from this scenario \citep{Hodges-Kluck2010B,Joshi2019}. 
Several XRGs with wings larger than the active lobes can be explained using the projection effect from the Back-flow model \citep{Giri2022A}; but, in a few instances the wings appear to be noticeably larger and complex than their active lobes \citep{Bruno2019,Joshi2019,Hardcastle2019,Ignesti2020}. 

In terms of emission signatures, such galaxies show distinctive spectral characteristics in low frequencies, where the geometrical wings exhibit sometimes flatter, steeper, or comparable 
spectral index ($\alpha$) distribution in comparison to the active lobes \citep{Lal2019,Gopal-Krishna2022}.
To explain this emission property, a dual AGN model was introduced \citep{Lal2019} where both the AGNs located at the centre of the same galaxy 
are hypothesised to expel jets at a certain angle, rising an X-morphology.

However, recent simulations and observational works strongly indicate that the diverted backflowing material can produce random shock sites in the wing that are able to 
re-energize the cooling particles, which most likely led to such a confounding property seen in wings 
\citep{Giri2022A,Gopal-Krishna2022,Mahatma2023}. 
Although the Back-flow model appears to be capable of answering many of the salient characteristics of such radio galaxies, it is difficult to 
comprehend, whether this is the universal model \citep[for a review see,][]{Gopal-Krishna2012}. 

In this regard, a comprehensive numerical investigation of such models has become absolutely essential in order to be able to determine how the differences appear in the formed XRGs, 
despite the employment of an identical initial configuration of the jet-ambient medium structure. 
Therefore, this science goal has become the central objective of the present study.

A subsequent part of this work would be to understand the jet-environment interaction of such peculiar radio galaxies.
This mechanism of interaction has been widely investigated through X-ray imaging studies of usual jetted sources. 
These studies conclude that as the jet propagates through the ambient medium (i.e., in a galaxy, a galaxy group, or a galaxy cluster environment), it creates an over-pressured 
cocoon removing the ambient material from their original location, and resulting in the formation of a cavity \citep{Gitti2010,Hlavacek-Larrondo2012,Shin2016,Vagshette2017}. 
The region just outside this cavity is surrounded by a shocked shell, and when viewed in X-ray, an X-ray depression 
region (the cavity) enclosed by a bright rim is commonly found. Hence, for such usual jetted sources, it is not surprising to find a pair of cavities
that originate from the flow of the bidirectional straight jet \citep{Hlavacek-Larrondo2015,Vagshette2019}. 

Recent findings with deep exposure maps, on the other hand, have revealed X-ray cavities that show a high degree of complexity. 
For example, \citet{Ubertosi2021} have identified two pairs of cavities that are located at similar distances from the cluster center and are aligned
perpendicular 
to each other \citep[see also][]{Bogdan2014}. 
An even more complex example can be found in the Perseus Galaxy Cluster \citep{Fabian2017}, which shows off-axis cavities, rims, and a geometrically complex distribution 
of pressure waves \citep[also see][for the RBS 797 galaxy cluster]{Ubertosi2022}. 
Other galaxy groups or clusters with such intricate cavity structures include, for example,
M84 \citep{Bambic2023}, NGC 5813 \citep{Randall2015}, A2052 \citep{Blanton2011}, NGC 5044 \citep{David2011}, Cygnus A \citep{smith2002},  MS0735.6+7421 \citep{McNamara2009}. 

A detailed observational analysis 
supplemented by predictions derived from a few available simulations, have indicated that the jets in each of these sources have most likely deflected from their expected straight 
path, as a result of a jet re-orientation into a new direction \citep{Falceta-Goncalves2010,Cielo2018,Lalakos2022}. 

These examples are indeed highly interesting cases, however, only a few of these sources have been fully examined so far. 
It is becoming clear that the discrepancies appearing in the jet morphology also imprint their signatures on the surrounding medium, as they are highly intertwined. 
Dedicated studies that help to understand these imprinted ambient signatures and use them for a possible differentiation between different formation processes of 
these sources are yet to be conducted. 

Overall, this discussion clearly necessitates a focused set of simulations, aimed at a specific class of sources which undergo different formation scenarios, but are embedded in
a similar jet-ambient configuration. 
Applying such a strategy in our present study, we aim to understand the physical mechanisms by which X-shaped radio galaxies disrupt their surrounding medium.
In particular, we will model the thermal free-free emission of the host cluster medium, that will allow us to further determine any potential distinction of different formation scenarios.

This paper is arranged as follows.
We describe the XRG formation models adopted for our simulations and their numerical implementation in Section~\ref{sec:Numerical Setup}. 
The dynamical evolution and characteristics of the evolving structures adopted in different 
runs are described in Section~\ref{Sec:Dynamical evolution: Development of the wings}. 
Section~\ref{Sec:Quantitative comparison}  explores the influence of these radio structures on the surrounding medium, while Section~\ref{Sec:Emission Result: Appearance of cavity regions} 
focuses on the examination of these signatures through X-ray mapping.
Section~\ref{Sec:Discussion} showcases a pertinent discussion on the impact of the viewing angle on the structure's appearance and age assessment.
We then summarize our work in Section~\ref{Sec:Summary}. 
In Appendix~\ref{Sec:Large-angle reorientation of Jet axis}, we provide a general exploration of potential jet-reorientation processes.

\section{Numerical Setup} \label{sec:Numerical Setup}
We performed a total of six simulations to better understand the formation and evolution of XRGs based on two different scenarios: the Back-flow case and the Jet re-orientation case. 

We applied the PLUTO code for our simulation study \citep{Mignone2007}. 
This code is tasked with solving a set of conservative relativistic magneto-hydrodynamic (RMHD) equations.
We define $D =\gamma\rho$ as the density in the observer frame, with the rest mass density $\rho$ and the Lorentz factor $\gamma$.
\textbf{\textit{v}} is the three velocity, and 
$b \equiv [b^0,\textbf{\textit{b}}] = [\gamma\textbf{\textit{v}}\cdot\textbf{\textit{B}}, 
\textbf{\textit{B}} \slash \gamma+\gamma(\textbf{\textit{v}}\cdot\textbf{\textit{B}})\textbf{\textit{v}}]$ 
is defined as the covariant magnetic field. We further define 
$w_t = \rho h + \textbf{\textit{B}}^2/\gamma^2 +\textbf{(\textit{v}}\cdot\textbf{\textit{B}})^2$ as the total enthalpy (with the specific enthalpy $h$), 
$\textbf{\textit{m}} = w_t\gamma^2\textbf{\textit{v}}-b^0\textbf{\textit{b}}$ as the momentum density, 
$P_t = P_g + \textbf{\textit{B}}^2/2\gamma^2 +\textbf{(\textit{v}}\cdot\textbf{\textit{B}})^2/2$ as the total pressure of the system (with the gas pressure $P_g$),
$E_t = w_t \gamma^2 - b^0 b^0 - P_t$ as the total energy,
and $I$ as the identity matrix, respectively. 
The conservative equations therefore can be written as
\begin{equation}
\frac{\partial}{\partial t}
    \begin{pmatrix}
    D\\\textbf{\textit{m}}\\E_t\\\textbf{\textit{B}}
    \end{pmatrix}
    + \nabla \cdot
    \begin{pmatrix}
    D\textbf{\textit{v}}\\
    w_t\gamma^2\textbf{\textit{v}}\textbf{\textit{v}}-\textbf{\textit{b}}\textbf{\textit{b}}+IP_t\\
    \textbf{\textit{m}}\\
    \textbf{\textit{vB}}-\textbf{\textit{Bv}}
    \end{pmatrix}^T
    =0,
\end{equation}
\citep{Mignone2009}.
The simulations were conducted in Cartesian geometry, hence the vectors (denoted by a bold font) represented in the above equation  
have three components along $(x,y,z)$.

These equations are solved in space to a second order of accuracy using linear reconstruction and the HLLC Riemann solver \citep{Mignone2006}. 
We ensure the solenoidal condition of magnetic field (i.e., $\nabla\cdot \textbf{\textit{B}}=0$) in the simulation domain by using the divergence 
cleaning approach described in \citet{Dedner2002}. 
We conducted the runs using the Taub-Matthews equation of state \citep{Taub1948, Mignone2005}.

All of our runs were performed using Adaptive Mesh Refinement (AMR) technique using the PLUTO code \citep{Mignone2012} over a 3D domain of size $400 \times 400 \times 400$ kpc$^3$. The simulation domain has a base grid of $96 \times 96 \times 96$, which using 3 levels of refinement results in an effective grid of $768 \times 768 \times 768$. This corresponds to a resolution of 520 pc.
Using an adaptive refinement criterion, we ensured that the jet injection region is always resolved at the highest refined level implying 5 grid cells per jet radius.

\subsection{Dynamical setup}
We concentrate on understanding the formation and long-term evolution of XRGs in a galaxy cluster environment. 
Existing observations of such radio galaxies have suggested that their host ambient media are found  to be ellipsoidal in nature \citep{Capetti2002,Kraft2005,Saripalli2009,Hodges-Kluck2010A,Hodges-Kluck2011}.
To initialise such a medium, we used the King's density profile which is defined as \citep{Cavaliere1976},
\begin{equation}
\rho = \rho_0 \left(1+\frac{x'^2}{a^2}+\frac{y'^2}{b^2}+\frac{z^2}{a^2}\right)^{-3/4}
\end{equation}
where, $\rho_0$ is the density of the cluster at the centre, set to 0.01 amu cm$^{-3}$, and $a$, $b$ are the effective core radius set to be 33 kpc and 50 kpc, respectively \citep{Heinz1998,Reynolds2002,Hardcastle2013}. 
This choice of $a$, $b$ provides a medium with ellipsoidal shape of ellipticity 0.3, where the major axis lies along the $y-$direction and the minor axes lies along the $x-$ and $z-$direction. Here, 
\begin{equation}
    x' = x {\rm cos}{\psi}- y{\rm sin}\psi,\ \ y' = x{\rm sin}\psi+y{\rm cos}\psi
\end{equation}
by which, we have rotated the ambient medium slightly from the Cartesian axes (rotated by an angle $\psi$ with respect to $z$-axis, with $\psi = 10^{\circ}$; see Fig.~\ref{fig:cartoon})
to break any symmetry in morphology that may appear while a jet flows along any symmetry axis (i.e., along major or minor axes) of such an ellipsoidally shaped ambient medium 
\citep{Rossi2017}. 
We describe the pressure distribution as following the density distribution, resulting in an isothermal atmosphere of 2 keV temperature. 

In order to keep this ambient medium in its initial static equilibrium, we apply a gravitational potential with a profile derived from the static fluid condition, $\nabla P_g = \rho \textbf{g}$ 
(with the gravitational acceleration $\textbf{g}$).  
We do not include any cooling of the cluster environment, as this study focuses on large scale jet induced morphologies
rather than the energetics of such inflated activities. 
Furthermore, the cooling time for cluster cores is typically of the order of a several hundreds of Myr and increases with distance from the center \citep{Hlavacek-Larrondo2012}. 
Compared to the average cooling time, our simulation operate for a shorter period of time.

Further, we inject a bidirectional jet into the computational domain using a cylindrical nozzle from the center of the cluster (introduced as an internal boundary). 
The radius of the jet injecting nozzle is 2.5 kpc ($r_j$), and its height extends over [-2.6, 2.6] kpc, 
which is divided into three zones. We denote them as the {\em neutral region} within [-1.0, 1.0] kpc, the {\em jet region} within [1.0, 2.6] kpc, and the {\em counter jet region} within [-2.6, -1.0] kpc, aiding us in injecting the bidirectional jet \citep[see][]{Rossi2017}. 

The jet is under-dense by a factor of $10^{-6}$ compared to the surrounding medium, $\rho_j = 10^{-6}\rho_0$ \citep{Rossi2017}, which is injected along the jet propagation direction with a bulk Lorentz factor of $\Gamma=3$ (the jet has no rotational velocity). 
It should be noted that the jet propagation direction changes with the different models considered in the present study, as discussed further in Section~\ref{Sec:Adopted Models}.  
We incorporate a toroidal magnetic field along with the jet flow (the average toroidal field strength is $B_t = 5.8\ \mu{\rm G}$) 
in the form as described in \citet{Rossi2017,Giri2022A}. 
The parameters defined within the jet injection zones are kept to be unchanged throughout the simulation time. 

The outside boundaries for these runs are established on all sides by extrapolating the ambient values. 
The above parameters
for the injected jet were also chosen primarily because of the statistically found radio power of XRGs, 
which lies in the \citet{Fanaroff1974} (FR) I/II boundary \citep{Saripalli2009,Landt2010,Gillone2016,Garofalo2020}.

In this regard, the power $Q_j$ for our
jet can be evaluated as follows \citep[using][subtracting rest mass energy]{Mignone2007A,Mukherjee2020},
\begin{equation}
    Q_j = \pi r_j^2 v_j \bigg[ \Gamma (\Gamma - 1)\rho_j c^2 + \frac{\rotatebox[origin=tr]{90}{$\prec$}}{\rotatebox[origin=tr]{90}{$\prec$} - 1}\Gamma^2 P_j + \frac{B_t^2}{4\pi} \bigg]
\end{equation}
where $c$ is the speed of light and 
$v_j$ is the jet bulk velocity represented through $\Gamma$, and
\rotatebox[origin=tr]{90}{$\prec$}, $P_j$ are the adiabatic index and pressure of the hot gas in jet, 
respectively. 
The adopted jet parameters further raise a jet power of $1.4 \times 10^{45}$ ${\rm erg\ s^{-1}}$ \citep[similar to][]{Giri2022A}.
The ratio of jet power to radio power at 1.4 GHz ranges from 3 to 300 as typically evaluated by \citet{Birzan2004}, resulting in a radio power estimate of $\leq 4.6\times10^{44}\ {\rm erg\ s^{-1}}$ for the XRGs; this value corresponds favorably with the power distribution of such winged radio sources.

We highlight that all of our simulations were conducted in code units, and their conversion back to physical scales is based on three basic units: 
$0.01$ ${\rm amu\ cm^{-3}}$ for density, 
$2.998\times 10^{10}$ ${\rm cm\ s^{-1}}$ for velocity ($c$), and 
$100$ ${\rm kpc}$ for length. 
All other units can be derived from these three. For example, the unit time is defined by comparing unit length to unit velocity and is $0.326\ {\rm Myr}$. 

Furthermore, we inject standard passive scalar tracers in the domain, ${\rm C_{jet}}$ and ${\rm C_{amb}}$ (values varying between $[0,1]$), representing jet and ambient medium, respectively. 

\subsection{Adopted Models}\label{Sec:Adopted Models}
\begin{figure*}
\centering
	\includegraphics[scale=0.62]{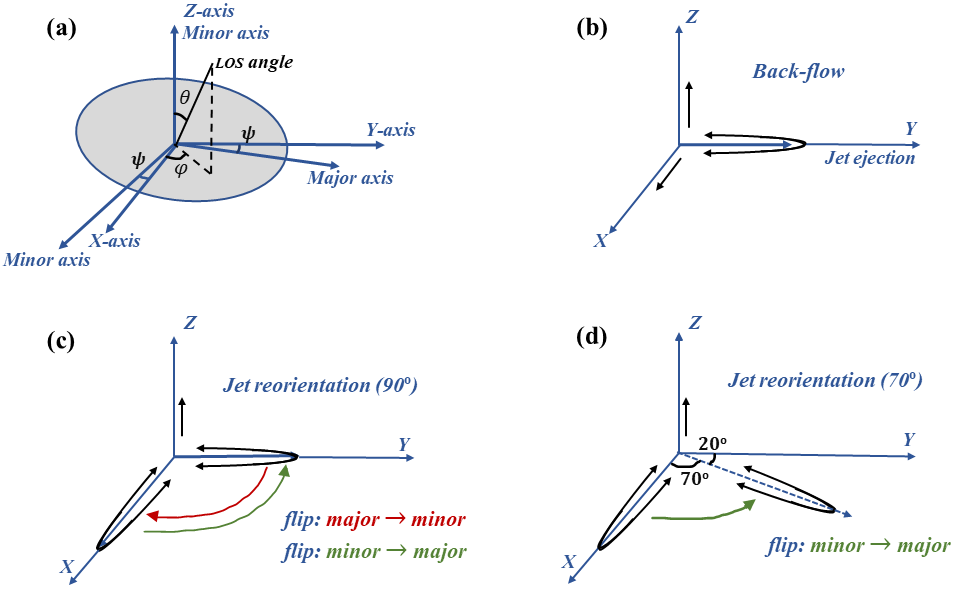}
    \caption{A schematic diagram representing the ambient medium (the elliptical shaded region)
    implemented in our simulations along with the line-of-sight (LOS) visualisation angle ($\theta, \phi$), and the adopted models for XRG formation. 
    The sub-figure (a) represents the ambient cluster medium configuration with respect to the Cartesian axes having size of $400 \ {\rm kpc^3}$. Sub-figure (b) depicts the Back-flow model, in which the jet only moves along the $y-$direction and the backflowing material moves in the other lateral directions. Sub-figure (c) showcases the Jet-reorientation scenario, in which the jet flips to an angle of $90^{\circ}$ from $x-$ (major) to $y-$ (minor) direction or the vice-versa, depending on the case under consideration. Sub-figure (d) represents the same situation as (c), but here the jet reorients to an angle of $70^{\circ}$. For visualisation simplicity, we confine the jet re-orientation activity to the $z = 0$ plane only.
    }
    \label{fig:cartoon}
\end{figure*}

Our first setup investigates the prediction of the Back-flow model of XRG formation. 
As investigation of a prediction,
we inject a bidirectional jet from the centre of the cluster medium that is ejected along the major axis of the medium (along the $y-$axis; Fig.~\ref{fig:cartoon}).
Jets propagating along the minor axis of such an ellipsoidal medium produce only a classical radio galaxy rather than an X-shaped Radio galaxy (as the model suggests) \citep{Capetti2002}. 
We run the simulation of this setup\footnote{The evolution time will be same for all other runs.} for 114 Myr, 
following the long-term evolution of the formed system. We denote this run as `bf{\small \texttt{00}}\rule{2mm}{.5pt}{\small YY}' (see Table~\ref{Tab:Parameters}). 

Our next set of simulations examine the Jet-reorientation scenario in the formation of XRGs, which are anticipated to arise from the spin axis flip of the jet ejecting SMBH \citep{Gergely2009,Roberts2015}. 
In this regard, we propose two scenarios for the timescale of the jet reorientation: 

\begin{enumerate}[label=(\alph*)]
\item quick reorientation `qr' \citep[changing the spin axis of the SMBH almost immediately, as suggested in several analytical exercises;][]{Merritt2002, Rottmann2002}, and 

\item slow reorientation `sr' \citep[changing the spin axis of the SMBH rather slowly as predicted in several numerical and observational results;][]{Zier2001,Dennett-Thorpe2002,Gong2011,Babul2013,Horton2020}. 
\end{enumerate}

Here, we have conducted simulations that take into account both of these possibilities (Fig.~\ref{fig:cartoon}), such that the injecting nozzle jet ejection direction is changed or flipped from its initial orientation to a specific angle 
either fast (immediately) or gradually \citep[5 Myr;][]{Dennett-Thorpe2002,Ubertosi2021}. 
While performing the simulations, the jet expulsion from the jet nozzle continues. 

Furthermore, we need to consider two sub-cases here, where the flip of the jet propagation direction happens from minor ($x$) to major ($y$) axis and major ($y$) to minor ($x$) axis of the cluster medium, respectively (see Fig.~\ref{fig:cartoon}). 
The above choice is made to naturally explain the statistically observed phenomenon of XRGs with wings aligning along the minor axis of the ambient medium (mostly observed) or along the major axis (several examples discovered recently) \citep[e.g.,][]{Joshi2019}. 
Hence, using this model, a total of four runs were performed, labelled as `qr{\small \texttt{90}}\rule{2mm}{.5pt}{\small XY}', `qr{\small \texttt{90}}\rule{2mm}{.5pt}{\small YX}', `sr{\small \texttt{90}}\rule{2mm}{.5pt}{\small XY}', and `sr{\small \texttt{90}}\rule{2mm}{.5pt}{\small YX}' (Table~\ref{Tab:Parameters}).
In each of these instances, the jet propagated in the same direction for 78 Myr before flipping to a certain angle, and then continuing to propagate upto 114 Myr \citep[see, e.g.,][]{Merritt2002,Bogdan2014}. 

It can be inferred from the above discussion that we shifted the jet ejection axis from its initial propagation direction to an angle $90^{\circ}$ in these cases.  
Although this scenario might seem distinctive, it is actually not an uncommon one. 
If the Jet-reorientation scenario is applicable, then the distribution of orientation angles between the active lobes and the wings provide indirect evidence in favour of a large-angle flip in the 
jet ejection axis for XRGs \citep{Capetti2002,Saripalli2009,Landt2010,Gillone2016,Joshi2019}.
Some more occurrences of a jet axis flipping by $\sim 90^{\circ}$ have been recorded, including \citet{Chon2012,Gitti2013,Bogdan2014,Ubertosi2021}, by assessing the direction of jet 
propagation and the jet-ambient medium interaction in multi-band observations. 
The flip of the SMBH spin axis by such a magnitude is supported also by a number of analytical and numerical investigations, such as done by \citet{Merritt2002,Lousto2015,Lousto2016,Lalakos2022}, 
who suggested a potential cause of such a process as either given by inhomogeneous gas accretion or by a binary black-hole merger (see Appendix~\ref{Sec:Large-angle reorientation of Jet axis} for further discussion).

Considering a spin axis flip by $90^{\circ}$ to be an extreme case, we ran another simulation with a jet propagation direction change to $70^{\circ}$ \citep[e.g.,][]{Capetti2002,Babul2013,Liu2019}. 
This additional simulation is motivated in order to see if a non-perpendicular flip ($<90^{\circ}$) can also develop the XRGs as observed, and to look for any possible signatures that
could disseminate between 
different formation scenarios and further constrain the involved parameters (labelled as `qr{\small \texttt{70}}\rule{2mm}{.5pt}{\small XY}'; Table~\ref{Tab:Parameters}). 
This reorientation, which we initiate by shifting the jet ejection direction from the x-axis towards the y-axis, is anticipated to result in a wing along the minor axis and an active jet within $20^{\circ}$ 
from the major axis of the ambient medium (Fig.~\ref{fig:cartoon}(d)). 

\subsection{Emission setup}\label{Sec:Emission setup}
To better understand such curved jetted inflated morphologies, we have additionally focused on modelling the thermal free-free emission of the simulated cluster medium in order to create a synthetic X-ray map of it. To separate the ambient material from the jet plasma, we used the passive tracer ${\rm C_{amb}}$ with ${\rm C_{amb} \geq 0.8}$. In this regard,
the emissivity of free-free (thermal Bremsstrahlung) emission can be expressed as:
\begin{equation}
    j_{\nu}(n, Z, T) = n_en_i\ \chi(Z,T)
\end{equation}
where, $n_e$, $n_i$ are the number density of electrons and ions in the gas
\citep[assumed to be equal;][]{Heinz1998}, and $\chi(Z,T)$ is a function of temperature $T$ and metallicity $Z$ \citep{Gronenschild1978}. 

Using the CHIANTI atomic data base, the function $\chi(Z,T)$ can be calculated and integrated over an energy band for a specific metallicity \citep{Landi1999}. 
By setting the metallicity at a level of 0.3 solar, we further evaluate $\chi$ in the X-ray (Chandra) band of [0.5-5] keV \citep{Hardcastle2013}
over a temperature range of $10^6$ - $10^{10}$ ${\rm K}$. 

\begin{figure}
\centering
	\includegraphics[scale=0.2]{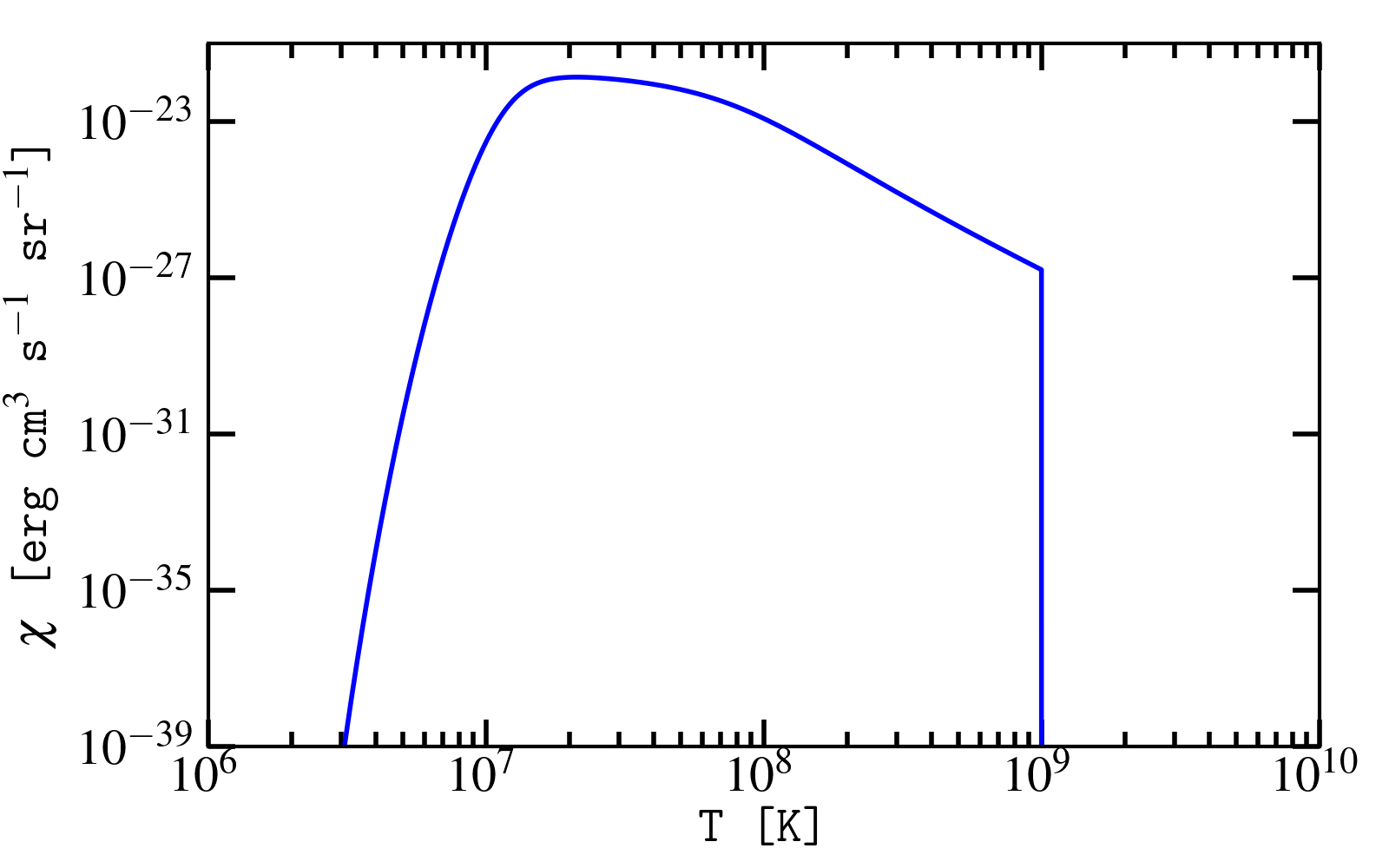}
    \caption{Distribution of the thermal emission coefficient $\chi$ over a range of temperatures $T$, as obtained for the chosen X-ray band of [0.5-5] keV and a metallicity of 0.3 solar.}
    \label{fig:emission_value}
\end{figure}

This choice of parameters yields a data set for the function $\chi$ including the corresponding temperatures. 
We show the data set in Fig.~\ref{fig:emission_value} within the designated temperature setting, displaying variations similar to those seen in \citet{Castellanos2015}. 
We further gather the number density and temperature (both weighted with tracer values) in each computational cell from the simulation results and use a linear interpolation to calculate the emissivity for the given X-ray energy band \citep[][]{Castellanos2015}. To obtain the intensity map, we further integrate the emissivity data-cube along a specific line of sight (LOS) angle ($\theta, \phi$) (Fig.~\ref{fig:cartoon}), under the assumption of an optically thin medium.

\begin{table}
\caption{Here, we highlight the simulation settings that are varied. In the first column, we denote the Simulation names, and in the rest columns, we represent the jet propagation direction change related parameters. Rest of the jet-ambient parameters are kept same for all the runs.}
\label{Tab:Parameters}
\centering
\begin{tabular}{ l|c|c|c } 
 \hline
 Sim. label& Flip & Flip & Flip\\
 & time (Myr) & direction & angle\\
 \hline
 bf{\small \texttt{00}}\rule{2mm}{.5pt}{\small YY}& $-$ & major ($y$) & $0^o$ \\ 
 qr{\small \texttt{90}}\rule{2mm}{.5pt}{\small XY}& Instantly & minor ($x$) $\rightarrow$ major ($y$) & $90^o$\\ 
 qr{\small \texttt{90}}\rule{2mm}{.5pt}{\small YX} & Instantly  & major ($y$) $\rightarrow$ minor ($x$) &$90^o$\\
 sr{\small \texttt{90}}\rule{2mm}{.5pt}{\small XY} & $5$  & minor ($x$) $\rightarrow$ major ($y$) &$90^o$\\
 sr{\small \texttt{90}}\rule{2mm}{.5pt}{\small YX}& 5  & major ($y$) $\rightarrow$ minor ($x$) &$90^o$\\
 qr{\small \texttt{70}}\rule{2mm}{.5pt}{\small XY}&  Instantly  & minor ($x$) $\rightarrow$ major ($y$) &$70^o$\\
 \hline
\end{tabular}
\end{table}

\section{Dynamical evolution: Development of the wings}\label{Sec:Dynamical evolution: Development of the wings}
Here, we discuss and compare dynamical aspects of simulations runs carried out in this study. 
We begin with a detailed analysis of a defined reference case, and then compare it to the dynamical properties of the other simulation runs.  

\begin{figure*}
\centering
	\includegraphics[scale=0.82]{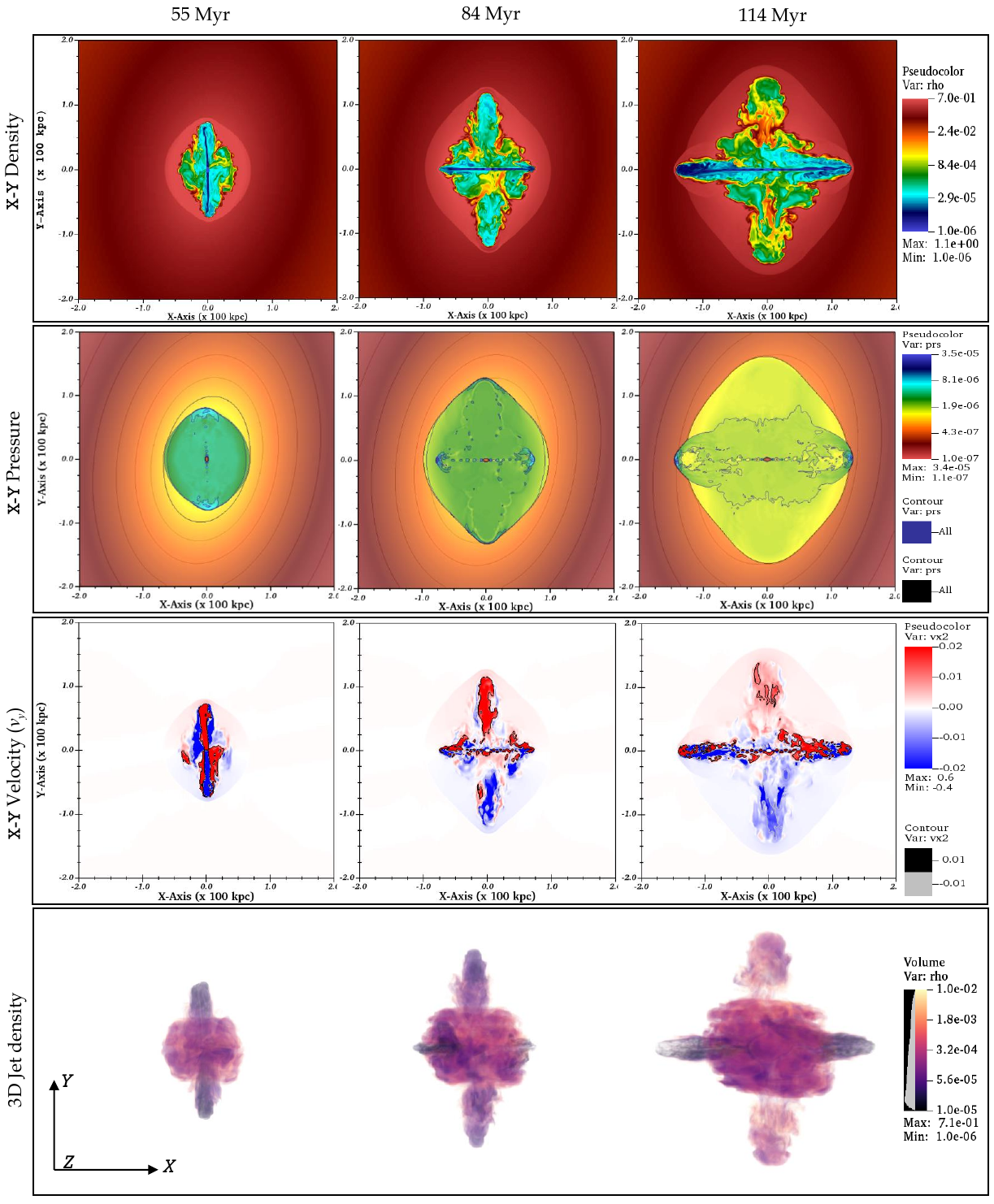}
    \caption{Representation of the evolution of a prominent XRG structure formed in the case `qr{\small \texttt{90}}\rule{2mm}{.5pt}{\small YX}' (fast Jet-reorientation scenario; Table~\ref{Tab:Parameters}). The top three rows represent $x-y$ slice of density, pressure and $y$-component velocity distribution of the evolving structure, respectively. The last row showcases the 3D volume rendered plot showing the intricate jetted structure as viewed along the $z$-axis. The corresponding time of evolution is highlighted at the top of each column. The colorbar for the diagrams is kept at the rightmost part and is the same for the associated row, presented in computational unit (multiply unit density ($0.01\ {\rm amu\ cm^{-3}}$), unit pressure ($1.5\times 10^{-5}\ {\rm dyn\ cm^{-2}}$) and velocity ($c$) for physical scales). See Section~\ref{Sec:Reference case run} for details.}
    \label{fig:Dynamical_evolution}
\end{figure*}

\subsection{Reference case}\label{Sec:Reference case run}
We consider the case `qr{\small \texttt{90}}\rule{2mm}{.5pt}{\small YX}' as our reference case (Table~\ref{Tab:Parameters}). This simulation run yields the most noticeable X-shaped morphology among all other cases considered in the present work (see Fig.~\ref{fig:Dynamical_evolution}). 

The jet here initially travelled along the major axis direction of the ambient cluster medium (i.e. along the $y$-axis) for 78 Myr, 
during which it evolved through interaction with the ambient gas, resulting in a substantial amount of backflowing material
\citep[causing by a sharp jump in entropy near the jet head region;][]{Cielo2017}. 
The system produces an extended cocoon structure as it is expanding into the ellipsoidally shaped ambient medium. 

At 78 Myr, the jet undergoes a $90^{\circ}$ flip in its propagation direction, resulting in a propagation along the minor axis (i.e along the $x$-axis).
This reorientation of jet generates a wing structure along the major axis of the ambient medium and an active lobe along the minor axis.
While the latter evolves, it gives rise to a morphology whose formation scenario has long been debated \citep{Hodges-Kluck2010B,Joshi2019}. 
It is still unclear how such structures can be naturally explained from the Back-flow model.

It is also interesting to note that at 114 Myr, the active lobe has grown to a similar extent as the wing ($\sim 140\ {\rm kpc}$; Fig.~\ref{fig:Dynamical_evolution}), which was developed
in about 36 Myr, which is significantly less than the wing evolution time. 
This shorter evolution time is the result of the reoriented jet beginning to propagate through a less dense cocoon region (established by backflowing material from the
earlier jet) that has already swept aside the heavy ambient gas. 
Also, the jet now moves along a path with a steeper pressure gradient of the surrounding medium, which facilitates jet propagation
along that direction.

These details concerning the structural evolution further provide evidence for the natural explanation of another debated characteristics of the XRGs.
That is the ratio of wing length to lobe length, which has been observed to be less than, comparable to, or even greater than unity \citep[see][]{Gower1982,Gopal-Krishna2012,Joshi2019}.
In case of the Jet-reorientation scenario, this ratio is determined rather by the evolutionary stage of the galaxy than by the complex interplay between an over-pressured cocoon 
scenario and the viewing direction, as it is expected by the Back-flow model.
Our simulation demonstrates how collimated the wing has remained since its formation. 
Despite the fact that the Back-flow model can also explain wing structure collimation \citep{Rossi2017,Cotton2020}, a combination of three of the above-mentioned properties of XRGs is most 
likely to support a Jet-reorientation origin of such radio galaxies.

The reorientation process of the jet further resulted in the insertion 
of multiple shock surfaces into the cluster medium. 
At time 114 Myr, the active jet inflates the strongest shock with a sonic Mach number of $\mathcal{M_S} \sim 2.8$, whereas the expanding wing head inserts a shock 
with $\mathcal{M_S} \sim 1.5$ (evaluated from the density jump at the shock surface using standard Rankine-Hugoniot condition).
As the wing head propagates, the related bow shock expands throughout the surrounding medium.
We find the Mach number for this shell-like shock surface to be approximately $\mathcal{M_S} \sim 1.5$.
 
The above result further supports the argument that the presence of multiple shock fronts associated with a jet in such ambient medium implies an older jet reorientation activity, 
similar to e.g., \citet{Ubertosi2022}. 
The presence of such differently aligned shocks also demonstrates the enhanced disruptive capability of jetted structures with bent morphologies,
strongly supporting the idea that such structures can be responsible for the isotropic heating of galaxy clusters, as has been reported in a number of recent studies \citep[e.g., see][]{Cielo2018}.

In Fig.~\ref{fig:Dynamical_evolution}, we also show the pressure distribution of the evolving structure. 
The figure highlights that both the formed cocoon and the shocked ambient material in the cocoon's immediate surroundings achieve a near pressure match. 
Hence, the system as a whole expands in the same way.
This extended glob (in yellowish green) that is seen, shows a volume that has an order of magnitude higher pressure than the surrounding cluster medium 
(which is represented by the elliptical isophotes), indicating a dynamic structure, 
which may explain how such objects can develop into massive XRGs on scales of hundreds of kpc. 

In this regard, we find the linear expansion speed of the wing (along $y$) towards the end of the simulation of about $\sim 900\ {\rm km/s}$, 
which is supersonic in comparison to the sound speed of the cluster environment ($707\ {\rm km/s}$). 

The region with the highest pressure always appears at the location where the jet head and ambient medium are in contact, and are eventually forming hotspots. 
Highly pressurised knots are also visible in the recollimation shock areas that have developed within the jet, which have also been noticed and investigated by \citet{Kraft2005}. 
The associated pressure decreases as the cocoon grows in size, but the active jet attempts to maintain over-pressure by inflating fresh material into the cocoon, as seen in the pressure map at 114 Myr, 
showing a contour of slightly higher pressure enclosing the current jet. 

This result of cocoon expansion is also evident from the distribution of ($y$-component of) velocity ($-$ve component in blue and $+$ve in red color). 
At the earlier time ($< 78\ {\rm Myr}$), the structure grows as expected from a FR II jet in an ellipsoidal medium \citep[similar to][]{Rossi2017}. 
Around 84 Myr, when the jet just flipped leaving a fresh wing along the major axis, mostly expands based on the older jet's injected material. 
After such injected material settles around 114 Myr, the backflowing material from the newer jet, diverting and entering the developed wing via the channel produced by the earlier jet, governs its evolution. 

The 3D volume rendered map of density (at 114 Myr) captures this channel of matter propagation into the wing region quite well, which appears as a neck-like structure in the wing region. 
This 3D density plot is produced in a way that captures the changing jetted morphology well, and further reveals how turbulent the growing structure is. 
This is also evident from the $x-y$ density slice, which shows that such turbulence is caused by the back-flow of matter when this becomes diverted in a lateral direction. 
Moreover, this structure exhibits a high level of Kelvin–Helmholtz instabilities (KHI), resulting in significant mixing of the surrounding material. 

The average Alfv\'enic Mach number of the cocoon ($\mathcal{M}_A = v/v_A;\ v_A=B/\sqrt{4\pi\rho}$) at time $114\ {\rm Myr}$ is 4.97 (super Alfv\'enic state), a regime enabling the KHIs to grow \citep{Acharya2021,Giri2022B}.
Such turbulent structure further generates sites of random shocks, therefore making these radio galaxies an ideal target for studying particle acceleration \citep[see e.g.,][]{Mukherjee2021}. 
Being turbulent, the over-pressure cocoon expands anisotropically, forming shell-like expanding surfaces and stripes, as reflected in the 3D density rendered plot \citep[see also,][]{Falceta-Goncalves2010A}.

\subsection{Comparing jets with backflow and slow reorientation}
\label{Sec:Comparison with other simulated systems}

Two other cases where similar morphological characteristics as Fig.~\ref{fig:Dynamical_evolution} is observed are `bf{\small \texttt{00}}\rule{2mm}{.5pt}{\small YY}' and `sr{\small \texttt{90}}\rule{2mm}{.5pt}{\small YX}', respectively (Table~\ref{Tab:Parameters}).

\begin{figure*}
\centering
	\includegraphics[scale=0.82]{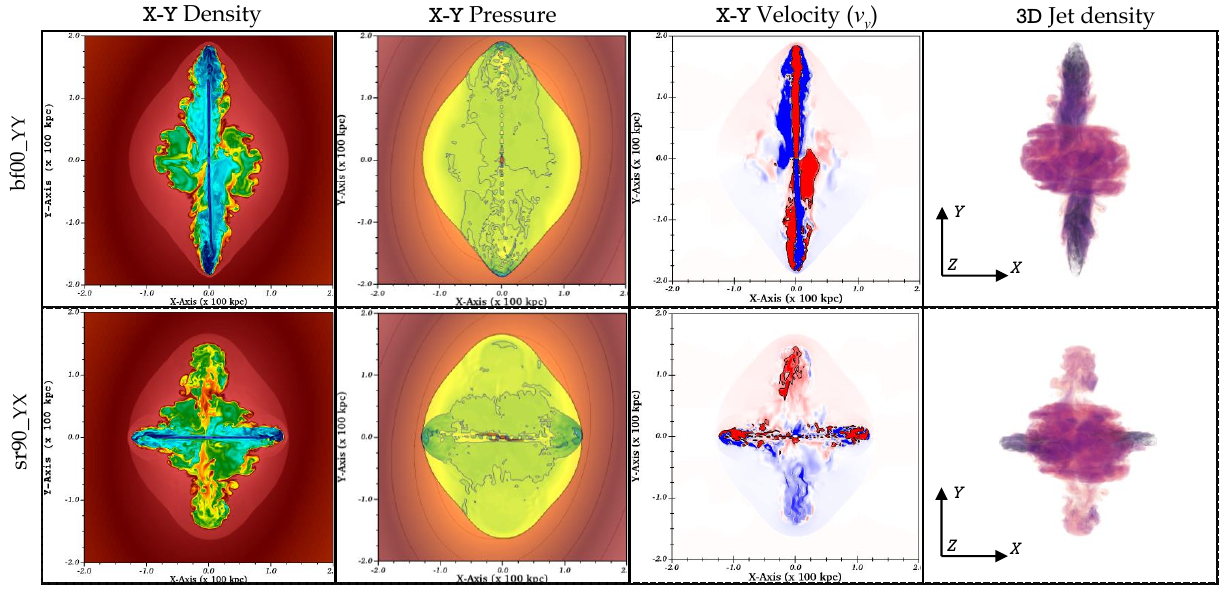}
    \caption{Representation of structures at 114 Myr that appear similar to the case `qr{\small \texttt{90}}\rule{2mm}{.5pt}{\small YX}' (Fig.~\ref{fig:Dynamical_evolution}). The associated simulation names are labeled in the leftmost part of the figure, referring to the Back-flow (top) and slow Jet-reorientation cases (bottom) (Table~\ref{Tab:Parameters}). The colorbar for the plots are kept same as in Fig.~\ref{fig:Dynamical_evolution}. See Section~\ref{Sec:Comparison with other simulated systems} for details and observational relevance.}
    \label{fig:Comparison_similar}
\end{figure*}

In the case of Back-flow scenario (Fig.~\ref{fig:Comparison_similar}; top panel), the jet only propagates along the $y-$direction (major axis of the ambient), resulting in a longer active jet length.
The backflowing material generated at the jet head propagates back towards the centre and deflects in the lateral direction, forming a wing-like structure \citep{Capetti2002}. 
The formed wing to active lobe length ratio in this case is $0.49$ ($< 0.8$; \citet{Cheung2007}), which struggles to qualify as a prominent wing structure. 
Rather, the structure can be identified as a classical radio galaxy with an off-axis lumpy extension \citep[][]{Chon2012}. 
To naturally explain XRGs with wing lengths greater than lobe lengths based on this model, \citet{Hodges-Kluck2011} have shown the need for an ellipsoidal ambient medium with even higher ellipticity.

However, such an ambient medium associating with an XRG is not always observed \citep[e.g., see][]{Hardcastle2019}.
Although the projection effects have an impact on the wing to active lobe length ratio \citep{Hodges-Kluck2011,Giri2022A}, it does not fully explain why some wings are much larger than the active propagating jet \citep[e.g.,][]{Bruno2019,Ignesti2020,Bruni2021}. 

We find that the wing expansion speed in this case is equal to the fast jet reorientation case `qr{\small \texttt{90}}\rule{2mm}{.5pt}{\small YX}' 
(presence of over-pressured cocoon is also seen here in the pressure plot; Fig.~\ref{fig:Comparison_similar}), demonstrating supersonic wing expansion. 
However, this expansion speed of wing is still $\sim2$ times less than the jet head propagation speed at 114 Myr. 

Additionally, we found that the wings do not appear well collimated here even when viewed along the $z-$axis. 
The contribution of wing to the observed structure will only increase if viewed from other viewing directions \citep{Giri2022A}.
Therefore, this model exhibits difficulties in generating wings with noticeable collimation, as opposed to the structure resulting from a jet reorientation. 
Nonetheless, XRGs with structural morphologies similar to the one shown in Fig.~\ref{fig:Comparison_similar} have been discovered, 
supporting the model's applicability \citep[e.g.,][]{Bruni2021,Mahatma2023}.

The structure formed in `sr{\small \texttt{90}}\rule{2mm}{.5pt}{\small YX}' case (Table~\ref{Tab:Parameters}) appears quite similar in comparison to the case of `qr{\small \texttt{90}}\rule{2mm}{.5pt}{\small YX}' (Fig.~\ref{fig:Comparison_similar}; bottom panel). 
The average wing length emerges to be identical in both the cases (reorientation event occurred at the same time), but because the jet reorientation took 5 Myr to complete in the former case,
the active lobe length appeared to be slightly less than the fast reorientation scenario (124 kpc versus 136 kpc). 
On the morphological ground, however, it becomes very difficult to disentangle the underlying formation process, indicating that a reorientation time of 5 Myr is still too quick to leave any 
observable signature over the formed XRG. 
In this regard, a longer jet reorientation time is thought to produce an S- or Z-shaped morphology rather than an X-shaped morphology \citep{Liu2019,Horton2020}.  
Hence, the parameter of reorientation time emerges as a critical parameter for understanding a wide range of winged sources, the investigation of which is beyond the scope of 
this paper and will be taken up in a separate work.

\subsection{Comparing jets reoriented from minor to major axis}

\begin{figure*}
\centering
	\includegraphics[scale=0.82]{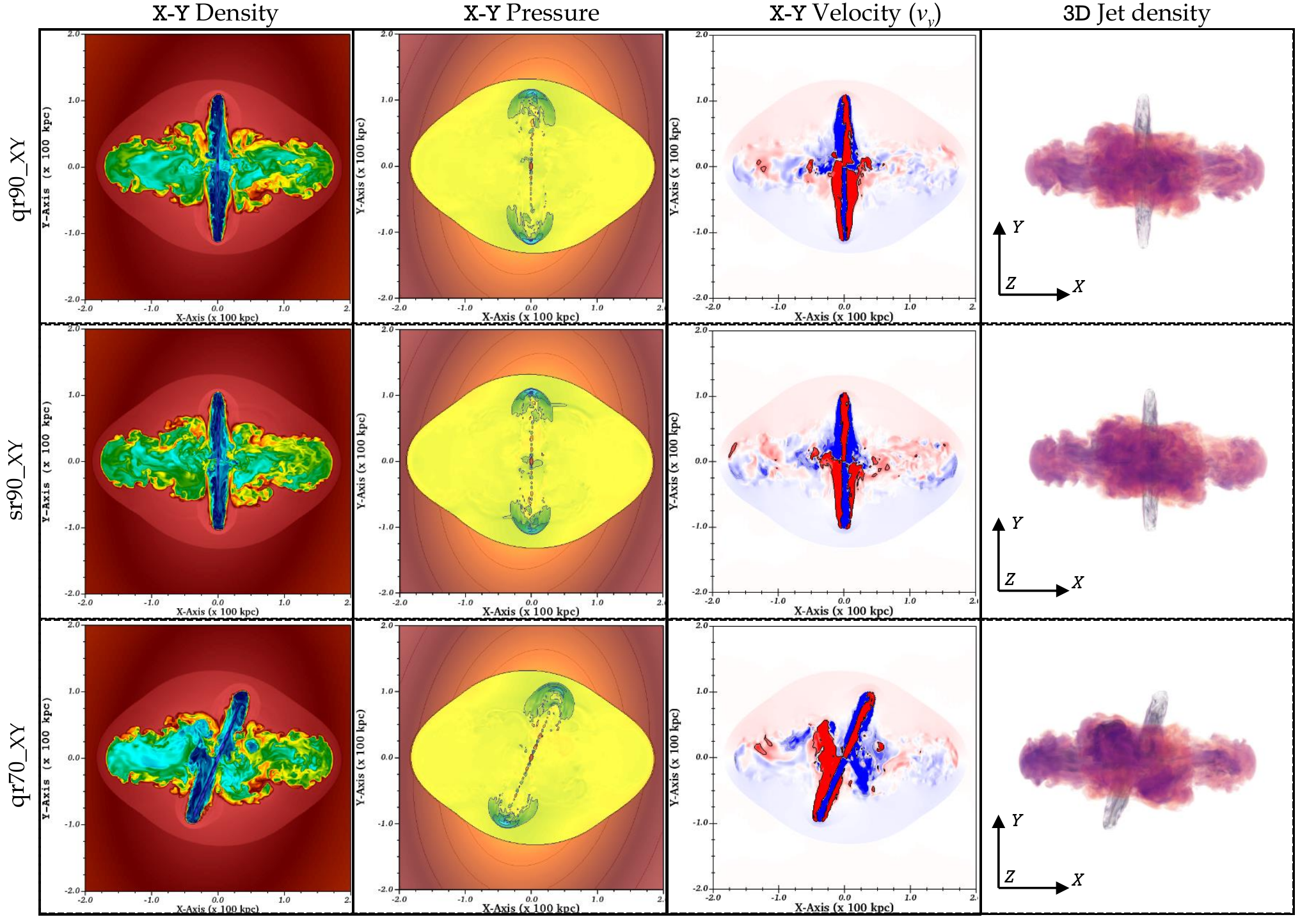}
    \caption{We demonstrate here three scenarios of a jet reorientation at 114 Myr, happening from the $x-$axis to the $y-$axis of the 3D configuration. The individual simulation labels are kept in the leftmost part of the diagram. All of these instances create an active lobe along the major axis and a pronounced wing along the minor axis of the ambient medium, typically observed in XRGs. The colorbar for each map is kept at the same level as in Fig.~\ref{fig:Dynamical_evolution}.
    See Section~\ref{Sec:Comparison with other simulated systems} for details and observational relevance.}
    \label{fig:Comparison_different}
\end{figure*}

We now discuss the dynamical aspects of three additional runs naming `qr{\small \texttt{90}}\rule{2mm}{.5pt}{\small XY}', `sr{\small \texttt{90}}\rule{2mm}{.5pt}{\small XY}' and `qr{\small \texttt{70}}\rule{2mm}{.5pt}{\small XY}' (Table.~\ref{Tab:Parameters}). 
In all of these cases, the jet reoriented from $x-$axis towards the $y-$axis, leaving  a prominent wing structure along the minor axis direction of the cluster medium (a salient characteristic of XRGs). 
The morphological appearance for all of these cases showcased in Fig.~\ref{fig:Comparison_different} at 114 Myr are unique, but not rare. 
Structures reported, for example, in \citet{Gower1982,Bruno2019} are reminiscent of the perpendicular flip case of the jet propagation axis depicted in Fig.~\ref{fig:Comparison_different}, 
while structures like observed by \citet{Murgia2001,Kraft2005,Ignesti2020,Gopal-Krishna2022} are reminiscent of the non-perpendicular flip scenario. 

The increasing number of discoveries of such sources poses  a formidable challenge to the existence of an universal model that will be able to explain 
the origin and evolution of all X-shaped radio galaxies. 

Among the three cases illustrated in Fig.~\ref{fig:Comparison_different}, the perpendicular Jet-reorientation cases exhibit a similar formed structure, indicating again that a jet flipping time of 5 Myr 
is still fast enough to impart a distinguishable signature that can be used to decipher the underlying rotation process.
The case of `qr{\small \texttt{70}}\rule{2mm}{.5pt}{\small XY}', however reveals some intriguing dynamical fingerprints on the developed structure. 
We found in this case the appearances of prominent yet complex filamentary structures, best captured in the 3D density map. 
The velocity ($v_y$) distribution (Fig.~\ref{fig:Comparison_different}) further emphasises the origin of such structures by attributing backflowing material from the newer jet that flows off-axis 
in comparison to both the wing and active lobe growth directions. 
Although not similar in morphology to the structure obtained here, prominent filamentary structures enclosing freshly propagating jet have been observed in XRGs \citep[e.g.,][]{Ignesti2020}.

The wing expansion speed is found to be high in all of the cases shown in Fig.~\ref{fig:Comparison_different} ($\sim 1050 {\rm\ km/s}$). 
This fast expansion is due to the propagation of the older jet and subsequently the current wing along the axis of highest pressure gradient in the ambient medium, facilitating the propagation.
The pressure gradient also eases the wing's lateral expansion,
producing a wing structure that is wider than its other counterparts, as illustrated in Fig.~\ref{fig:Dynamical_evolution}, and \ref{fig:Comparison_similar}. 
Such intricate structures inflate complex pressure waves into the cluster medium, which can further be identified in the pressure distribution map \citep[see also][]{Hardcastle2013,Cielo2018}. 
In each of these examples, the ratio of wing to active lobe length appears to be greater than $ 1.5$. 

The bow shocks that the wings are associated with (in Fig.~\ref{fig:Comparison_different}) have a sonic Mach number $\mathcal{M_S} \sim 1.9$, while for the  active lobes we find $\mathcal{M_S} \sim 1.6$.
The Mach number appearing with the active jet is smaller primarily because the lobe propagates in a shocked ambient medium resulted from the previous jet's propagation.

\section{Dynamical impact on the ambient medium} \label{Sec:Quantitative comparison}
In this section, we quantify the jet induced morphologies and their subsequent impact on the surrounding medium.

\begin{figure*}
\centering
	\includegraphics[scale=0.23]{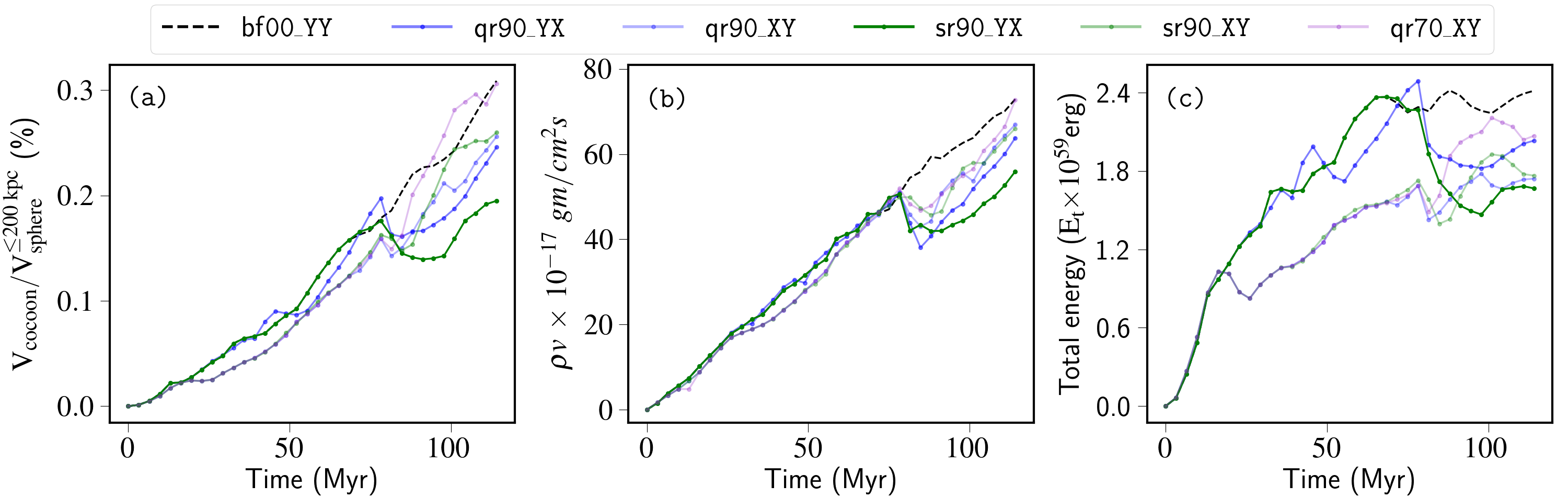}
    \caption{Quantitative representation of the structural evolution formed through various models (labelled in the top). The subplot-\texttt{(a)} depicts volume evolution of the growing cocoon in relation to a volume of sphere 200 kpc, demonstrating how much of the ambient cluster is perturbed. Subplot-\texttt{(b)} and -\texttt{(c)} show the mass momentum and total energy evolution of the same emerging structure. All of the images near 78 Myr show a slightly distinct evolutionary tendency in reorientation scenarios due to a considerable turbulent mixing of jet-ambient material (see Section~\ref{Sec:Quantitative comparison} for details).}
    \label{fig:quantitative}
\end{figure*}

The evolution of cocoon volume ${\rm V}_{\rm cocoon}$ inflated by the jet is represented in Fig.~\ref{fig:quantitative}(a). 
We present this with respect to the volume of a sphere of radius 200 kpc, denoted by ${\rm V}_{\rm sphere}^{\leq {\rm \huge 200\ kpc}}$, 
to illustrate how much of volume of 
the ambient cluster medium is perturbed.
We find that starting off with the same value, all of the structures emerging from different models evolve similarly in their early stages of evolution.
The Back-flow model follows this evolution beyond 78 Myr, with the straight jet and backflowing material's propagation, significantly disrupting the circumjacent medium at 114 Myr. 
This disruption is caused by the fact that the energy and momentum of the jet, as well as that of the backflowing material, are always channeled into the same specific direction. 

In difference, for the scenario of a perpendicular jet reorientation, we see a considerable volume decrement immediately after 78 Myr, thus just 
after the moment of the reorientation of the jet. 
At that point in time, the redirected jet travels through the already existing cocoon, while KH instabilities develop along the cocoon-cluster boundary, resulting in turbulent mixing of jet material 
(as discussed earlier).
During this redirecting activity of the jet, the inserted energy and momentum begin to be distributed over a broader region, thus having a weaker effect on the  expansion of the cocoon \citep{Cielo2018}. 
Then, after a brief period, the fresh jet begins excavating the cluster medium in the new direction of propagation, thereby increasing the cocoon volume again. 
Nevertheless, due to the two physical processes discussed above, it never matches the Back-flow scenario. 

The model applying a $70^{\circ}$ jet reorientation reveals a slightly different narrative by perturbing a similar volume of the cluster as like in the Back-flow case. 
We found that the jet-ambient material mixing through the KHIs is less effective here ($\mathcal{M_A}\sim 3.5$), leading to such a result.

A similar evolution is also reflected in the momentum ($\rho v$, thus absolute value of momentum) chart in Fig.~\ref{fig:quantitative}(b), 
which we plotted to understand the inflated disruptions caused by the evolving structures onto the cluster medium and on itself.

Jet and counter-jet are injected into the computational domain with zero net momentum. 
We thus plot the absolute value of the momentum, which is expected to grow with the jet inflation, further imparting its effect 
on the surroundings.
The absolute momentum measured at a time, is expected to be equal to the injected jet momentum until that time.

We have detected analogous evolutionary trends to those depicted in Fig.~\ref{fig:quantitative}(a) in the momentum magnitudes as well. 
This obtained evolution further suggests that the Jet-reorientation scenarios exhibit lower turbulence levels in terms of momentum values, 
with the exception of the $70^{\circ}$ jet-flip case, which is consistent with the Back-flow scenario. 
To better scrutinize this intrinsic dynamical properties of the formed structures from different models, we tracked the evolution of various 
momentum components that are expected to be associated with the jet evolution mechanisms. 

\begin{figure*}
\centering
	\includegraphics[scale=0.23]{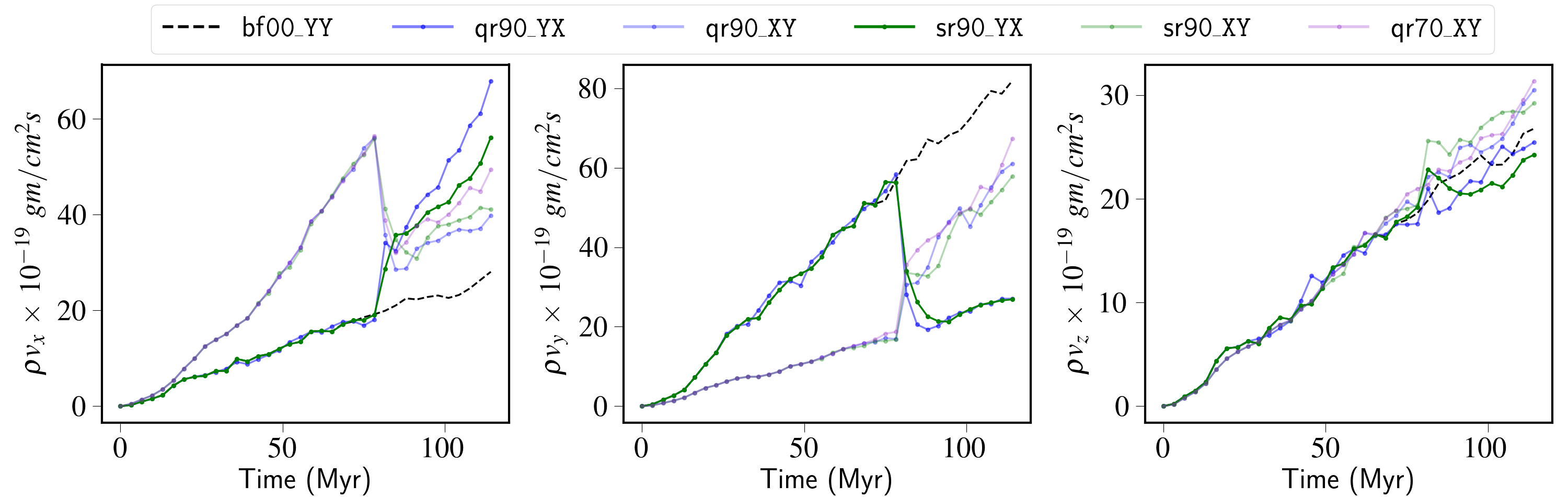}
    \caption{Evolution of different components of momentum ($\rho v_x, \rho v_y, \rho v_z$; absolute values), showcasing the dependence of quantitative variation on different jet evolutionary processes (or formation models). See Section~\ref{Sec:Quantitative comparison} for details.}
    \label{fig:momentum}
\end{figure*}

In Fig.~\ref{fig:momentum}, we show these individual components, $\rho v_x, \rho v_y, \rho v_z$. 
The trajectory of the momentum growth for the Back-flow model is characterized by a straightforward evolution throughout all plots. 
The dominant momentum component $\rho v_y$ here is generated by the active jet, which is imparted along the $y$-axis. 
Conversely, the back-flowing plasma contributes only a minor strength of momentum along the $x-$ and $z-$directions, thereby aiding in
the growth of the structure.

The evolution of the different Jet-reorientation scenarios differs in the magnitude and evolution of $\rho v_x$ and $\rho v_y$, but follows 
a similar trend in $\rho v_z$.
This identical trend in the momentum component along $z$-direction for all the cases arises from the fact that the structures along this axis 
grow only by the back-flowing plasma from the active jet head that propagates either along the $x-$ or $y-$axis. 
For the cases of a jet reorientation, we observe a bifurcation in the $\rho v_x$ and $\rho v_y$ plot (see Fig.~\ref{fig:momentum}). 
This may be actually expected, given that there exist two distinct subcategories of Jet-reorientation. 
These subcategories correspond to the jet-flipping from the minor axis ($x$) to the major axis ($y$) of the ambient, and vice versa.

To summarize, the distinct evolution patterns of individual momentum components of the cocoon plasma seem to depend on the particular XRG formation model. 
The observed divergence in course of evolution hence suggests that the magnitude of turbulence generated within a cocoon may vary across the different models. 
Consequently, we anticipate that varying degrees of particle re-energization will occur within the resultant XRGs, potentially resulting 
in discernible effects on the non-thermal emission maps.
This potentially interesting kinematic feature requires further investigation and verification, but is presently outside the purview of 
the present work, which has not particularly focused towards non-thermal emissions. 
Nonetheless, it should be pursued as a prospective area of investigation in the future.

The next quantity we discuss is the total energy ${\rm E_t}$ stored in the jet-inflated bubbles (parameters 
denoted by a subscript `b'), considering the kinetic, thermal and magnetic parts as \citep[follwing,][]{Cielo2018},
\begin{equation}
    {\rm E_t} = {\rm V}_b\bigg[\frac{1}{2}\rho_b v_b^2 + \frac{1}{\rotatebox[origin=tr]{90}{$\prec$} - 1}P_b+\frac{B_b^2}{8\pi}\bigg]
\end{equation}
pertaining to a sub-relativistic bubble expansion in which $\rotatebox[origin=tr]{90}{$\prec$}$ is $4/3$, given that the bubble constituents are relativistic. 

All of our runs exhibit the same initial evolutionary phase (upto $\sim 20$ Myr; Fig.~\ref{fig:quantitative}(c)), occurring in the
denser central region of the cluster. 
After this time, the evolutionary path split into two branches, with one path associating the cases with initial jet propagation along $y-$axis 
and the other path associating the cases with initial jet propagation along $x-$axis. 
Around 78 Myr, the reorienting jets exhibit the same trend as described earlier for the volume and momentum plot, shown in Fig.~\ref{fig:quantitative}(a) and (b), 
while the Back-flow scenario maintains the evolution. 

Regardless of the evolutionary phases, the values of ${\rm E_t}$ obtained at 114 Myr for all our model approaches are not substantially different from one another.
The evolutionary trend indicates that they all reach a steady state. 
For our simulation runs, the contribution to ${\rm E_t}$ from the kinetic part is $<4\%$, while from the magnetic part it is $<10\%$. 
The fractions we compute for the magnetic and kinetic energy components in regard to the total energy are similar in all cases.
The major contribution to ${\rm E_t}$ comes from the thermal part, which consists of the jet-inflated bubble's pressure ($P_b$) and volume ($V_b$). 

As time progresses, the volume $V_b$ increases, but at the same time the pressure $P_b$ decreases. 
As the structure undergoes expansion, the presence of instabilities and the mechanical work ($PV$) generated 
by the jet-induced cocoon
lead to the discharge of material and energy into the surrounding cluster medium.
As a result, these energy loss processes and the constant energy injection from the jet, gradually 
bring the system to a steady state.
However, at this point we cannot answer the question whether this stage will be maintained or not for future times.

We calculate the cumulative energy input from the jet within 114 Myr to be $5.03 \times 10^{60}$ ${\rm erg}$. 
The range of energy contained within the cocoon bubble at 114 Myr, as obtained for the different models, is between [$0.78 - 1.22$] $\times 10^{59}$ erg.
Note that a significant amount of energy has been transferred to the surrounding cluster environment, which is expected to heat up the cluster medium.
In addition, we notice that the jets performing a reorientation affect the ambient medium greatly by transferring a large amount of energy isotropically into the cluster medium, 
in difference to the Back-flow scenario.

\section{Emission Results: Appearance of X-ray cavities}
\label{Sec:Emission Result: Appearance of cavity regions}

In Section~\ref{Sec:Dynamical evolution: Development of the wings}, we demonstrated the evolution of over-pressured cocoons, and their dissemination creating shock surfaces in the circumjacent medium. Here, we express the effect of such activities on the ambient medium by generating the thermal X-ray emission maps as discussed in Section~\ref{Sec:Emission setup}.

\subsection{Cavity evolution: Reference case}\label{Sec:Cavity evolution: Reference case}
\begin{figure*}
\centering
	\includegraphics[scale=0.24]{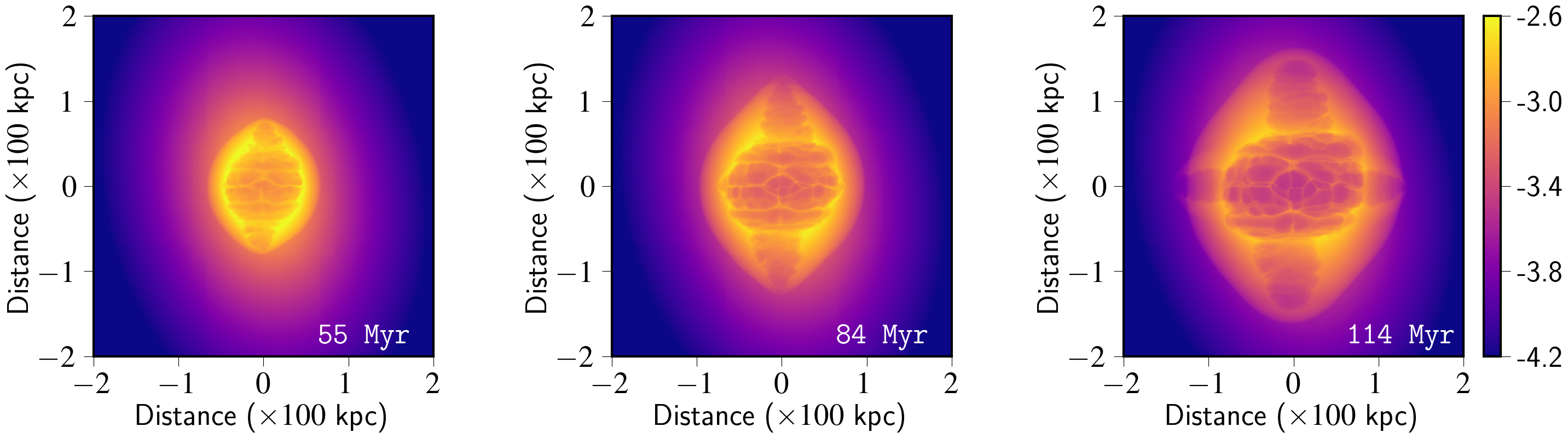}
    \caption{Thermal X-ray intensity map (in ${\rm log  [erg\ s^{-1} cm^{-2} sr^{-1}]}$) in the [0.5-5] keV band for model `qr{\small \texttt{90}}\rule{2mm}{.5pt}{\small YX}'.
    The sequence showcases the formation and evolution of X-ray dip regions (cavities) in the midst of the cluster. 
    The evolution is complementary to the cocoon geometry (Fig.~\ref{fig:Dynamical_evolution}, viewing angle ($0^{\circ},0^{\circ}$)). 
    Four distinct elongated cavities are generated, associated with wings and active lobes at 114 Myr, bearing signatures of the reorientation event. 
    See Section~\ref{Sec:Cavity evolution: Reference case} for details.
    }
    \label{fig:Cavity_evolution}
\end{figure*}

X-ray cavities (depression regions in X-ray map), are formed when the ambient cluster material is pushed out of its original place by the expanding over-pressured cocoon generated by an active jet. In Fig.~\ref{fig:Cavity_evolution}, the formation and evolution of a geometrically complex X-ray cavity system are shown for the case `qr{\small \texttt{90}}\rule{2mm}{.5pt}{\small YX}'. 
We see that the evolving cavity  behaves geometrically equivalent to the 3D density plot in Fig.~\ref{fig:Dynamical_evolution}. This further highlights the interconnected nature of the jet and surrounding environment.

At 55 Myr, the cavity region that has formed in the center of the cluster, is surrounded by a thick emission enhanced region (X-ray bright rim).
This structural evolution results from the strong shock that the cluster material has experienced 
because of the expanding, over-pressured cocoon 
(resulting inner brightest layer in the immediate surroundings of the expanding cocoon), 
as well as due to the bow shock from active jets 
(resulting outer surface of the thick shock zone). 
At this time, the jetted structure has only disrupted the central cluster regions, and hence the outer unperturbed material's emission, well captured in Fig.~\ref{fig:Cavity_evolution} (left), shows the maintained elliptical geometry of the cluster.

At 84 Myr upon re-orientation, the remnants of the older jet essentially results in an expanding wing producing elongated and collimated cavity structure in the X-ray map. 
The signature of such an actively expanding wing in terms of X-ray cavities is similar to that observed by \citet{Hlavacek-Larrondo2012,Randall2015,Shin2016} for an elongated bi-directional actively evolving jet. The newly formed reoriented jet also drills the cluster medium in the perpendicular direction, producing signatures of an additional pair of cavities along this latest direction of propagation.
This young emerging cavity pair associating the active jet head can be observed along the minor axis direction of the ambient medium in Fig.~\ref{fig:Cavity_evolution} (middle panel).

At 114 Myr, we clearly see the signature of four elongated cavities, with two of them aligning along the direction of the minor axis of the cluster,
associated with the actively evolving 
lobes. Whereas, the other two are less collimated but still elongated and aligned along the wings (Fig.~\ref{fig:Cavity_evolution}). 
At this stage, the cavities begin to appear prominently due to their increasing size and sharp contrast.
Quantitatively, the ratio of intensity between  the most prominent cavity and the surrounding bright shocked rim is 0.3.
Overall, the X-ray intensity varies between $0.002$ to $2.9 \times 10^{-5}$ ${\rm erg\ s^{-1} cm^{-2} sr^{-1}}$ (see the right panel of Fig.~\ref{fig:Cavity_evolution}), and the estimated X-ray luminosity $L_X \sim 10^{45}$ ${\rm erg\ s^{-1}}$. 
Given the active nature of the structure (still in an over-pressured phase), its further growth is anticipated.

From an observational standpoint, discovering four cavities connected to such events of a jet reorientation is not unusual. Examples are 
the cavity system discovered in the deep exposure X-ray map of the cluster RBS 797 \citep{Ubertosi2021}, or Cygnus A and NGC 193  \citep{Chon2012,Bogdan2014}. 
In the context of X-shaped radio galaxies, the observational work of \citet{Hodges-Kluck2010B} and the simulation work of \citet{Lalakos2022} have shown some support for such a 
conclusion of forming four cavities associating with jet reorientation activity.

During this time of 114 Myr, a new pair of bow shocks is produced by the reoriented jet in the cluster medium, 
creating X-ray bright rim regions that enclose the associated cavities. 
This generation of multiple shock surfaces reiterates the prediction made in Section~\ref{Sec:Reference case run} that such bent-jetted structures appear to have a greater 
disruptive effect on the ambient medium (compared to normal radio galaxies) and hence may be responsible for the isotropic heating of the ambient medium \citep[see][]{Cielo2018}. 

The cavity blob observed in the central region of the formed structure, associated with the fat cocoon, is also another noteworthy feature (Fig.~\ref{fig:Cavity_evolution}). 
The over-pressured extended cocoon bubble in the midst of the cluster medium (Fig.~\ref{fig:Dynamical_evolution}) at 114 Myr still evolves from the back-flowing material of the active jet and shows presence of X-ray depression region. 
This cavity blob is rather large in size and is surrounded on all sides by a shocked rim of increased emission. 
From a theoretical standpoint, the formation of such an extended structure is not unusual because 
of the evolving cocoon pushing aside the cluster material, as shown by \citet{Heinz1998} \citep[see also,][]{Bogdan2014}. 
In addition to this, we observe an imprint of filamentary stripes over this central X-ray cavity zone. 
These X-ray bright stripes appear because of the turbulent supersonic expansion 
of the cocoon, which we have also outlined in Section~\ref{Sec:Reference case run} \citep[see][]{Cielo2018}. 
The observed stripes are indicative of saddle lines that have formed between adjacent cocoon blobs that undergo supersonic expansion.

\subsection{Cavity identification}\label{Sec:Cavity quantification}
\begin{figure*}
\centering
	\includegraphics[scale=0.3]{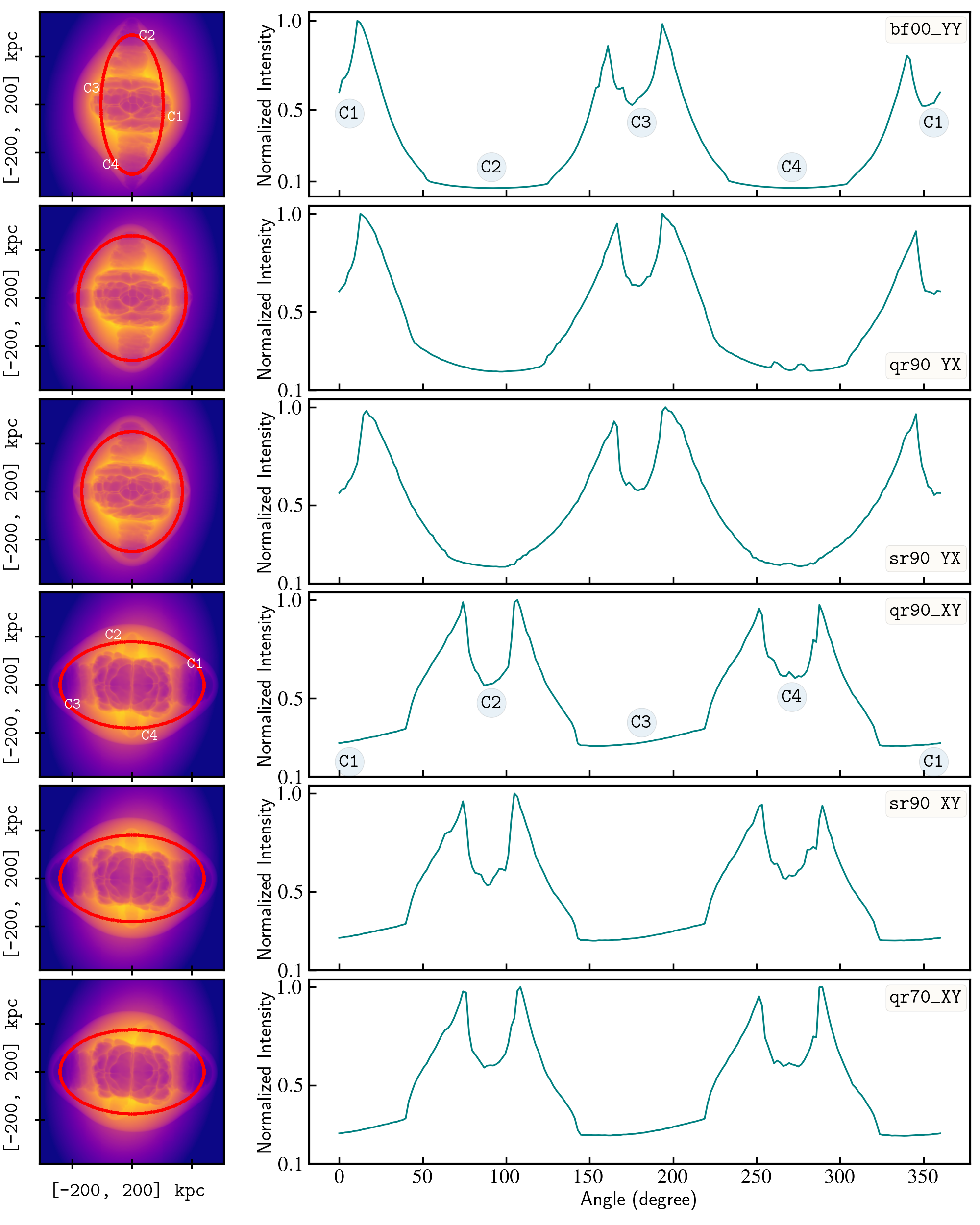}
    \caption{\textit{Left column:} Thermal X-ray intensity maps for six different cases that we have considered in Table~\ref{Tab:Parameters}, obtained at 114 Myr, plotted in the same colorbar interval as in Fig.~\ref{fig:Cavity_evolution} in ${\rm log[erg\ s^{-1} cm^{-2} sr^{-1}]}$. \textit{Right column:} Chart of normalized intensity variation versus azimuthal angle, as obtained by following an elliptical isophote over the X-ray intensity map in the immediate left, showcasing the detection and strength of X-ray cavities. The simulation names for each row are attached in the rightmost corner. To guide the eye in locating the position of X-ray cavities, we have labelled cavity names \texttt{C1, C2, C3, C4} in two of the above cases. All of these intensity maps are viewed along the $z$-axis, i.e., along the $(0^{\circ},0^{\circ})$ viewing angle.}
    \label{fig:Cavity_quantification}
\end{figure*}

The distribution of X-ray depression regions, respectively cavities, obtained for all the formation models is 
shown in Fig.~\ref{fig:Cavity_quantification}. 
The intensity mappings showcase the formation of a geometrically complex alignment of the X-ray depression regions in the cluster medium,
which is associated with the bent jetted structures shown in Fig.~\ref{fig:Dynamical_evolution},~\ref{fig:Comparison_similar},~\ref{fig:Comparison_different}.

We find that all cavity regions are enclosed by an outer bright rim of emission. 
For the models applying a jet reorientation, we see, in addition, that a new set of shock waves has been inserted into the cluster medium, in contrast to the Back-flow scenario. 
Furthermore, for the cases of a jet reorientation, we observe that the outer cavities which are 
connected to both the active lobe and wing, exhibit elongated morphology {\em along} the jet flow.
Conversely, in the Back-flow scenario, the cavity structure associated with the wing is elongated in a direction {\em perpendicular} to the growth of the wing. 

The investigation of elongated X-ray cavity structures, which are oriented either along or perpendicular to the jet propagation axis, was explored by \citet{Guo2015} through numerical simulations, and by \citet{Shin2016} through observations. Their findings help us understand that in the Back-flow scenario, the actively propagating jet, lighter in density and faster in propagation, will form an extended cavity along its path. In contrast, the expanding back-flowing plasma with slower expansion and heavier matter in wing (see Fig.~\ref{fig:Comparison_similar}) is expected to expand in the lateral direction to the wing growth direction, forming an elongated cavity that appears perpendicular to the growth path. 

In the Jet-reorientation cases, the exact same jet travels in each direction, relating the wing and lobe, resulting in radially extended cavities, i.e. alignment along the jet propagation \citep[e.g., see][]{Hodges-Kluck2012}.
Regardless of the distinct feature between cavity morphologies derived from Back-flow and Jet-reorientation models, it remains uncertain whether this could always serve as a viable distinguishing characteristic between the models. 
This uncertainty is due to the fact that the collimation of the wing in the Back-flow model is contingent upon the shape of the ambient medium and the direction of jet propagation \citep{Hodges-Kluck2011,Rossi2017,Cotton2020}. Additionally, a study conducted by \citet{Chon2012} has also demonstrated that the off-axis extended cavity in Cyg A, which is similar to that observed in the Back-flow case, can be formed from a scenario involving a perpendicular reorientation of the jet.

We have further investigated the azimuthal intensity distribution by using an elliptical isophote (shown in red line color in Fig.~\ref{fig:Cavity_quantification}).
This allows us to gain a {\em quantitative} understanding of the cavities formed, and to measure their extension.
In order to highlight the position of the X-ray depression zones, we have labelled the cavity locations with \texttt{C1, C2, C3, C4} in two of the cases in Fig.~\ref{fig:Cavity_quantification}. 
The isophotes are adjusted to navigate through the apex of the cavities due to their pronounced appearance at that location as well as to prevent potential overlap of structures that may 
occur with smaller isophotes.
In the context of comprehending cavity strength, \citet{Hlavacek-Larrondo2012} have categorized cavities into two classes based on the fractional difference between the emission 
originating from the X-ray dip area and its adjacent vicinity. 
The classification is based on the sample taken from the Chandra telescope's \textit{Massive Cluster Survey}.
The two classes of cavities are: {\em a) Clear cavities:} Fractional change $>$ 20\%,  and {\em b) Potential cavities:} Fractional change $<$ 20\%.

In this work, the cavities associated with the wing and lobe formed from the adopted models show fractional differences $> 35\%$. 
Hence, the X-ray depression zones can be referred to as {\em clear cavities}. 
In detail, we find that the strength (in terms of fractional difference) of the X-ray depression zones varies from 36 to 45\% for the active lobes and 70 to 81\% for the 
wings for all Jet-reorientation models. Our synthetic maps have not undergone any filtering procedures to imitate the resolution of a telescope, nor have we taken into account the effect of galactic absorption. Therefore, these fractional decrement values can be considered upper limits for the cavities detected here.
Nevertheless, these fractional decrement values we show above are in line with that estimated for the X-ray cavities seen in the Perseus A galaxy cluster \citep{Heinz1998}. Further, in our reoriented jet models, we observed that the cavities formed due to wings are more prominent as compared to the cavities produced by the active lobes. The strong shocks due to the reoriented jet forming the active lobes essentially contribute in reducing the contrast for the cavities observed for a given line of sight \citep{Cielo2018}.

In the context of the Back-flow model, we observe an alternative narrative. 
The actively propagating jet creates a lobe, corresponding to a cavity of greater contrast compared to the cavity 
associated with the wing. 
This phenomenon is not unusual and is documented in several numerical and observational investigations \citep{Wise2007,McNamara2009,Cielo2018}. 
The principal cause of such a formation of cavities by active lobes is the continuous jet propagation in a specific region, channeling matter and energy along one axis. This process allows the jet to surpass the central dense cluster area, thereby facilitating the expansion of the cavity.   

It is worth noting that the selection of an elliptical shaped isophote in order to identify the cavities does not introduce any bias in the assessment 
of the cavities. 
This statement is further supported by recent findings from \citet{Ubertosi2021}, which demonstrate that using a different shape, for example a rectangle that encompasses the apex 
of the cavity regions, would have minimal impact on the identification.

In summary, our findings indicate that an analysis of the cavity strength and shape distribution in the active lobes and wings may serve as a method of distinguishing between the 
Back-flow and Jet-reorientation models.

\section{Discussion}\label{Sec:Discussion}
This section discusses two crucial issues: the appearance of cavity geometry under various viewing angles, and the assessment of dynamical age considering various temporal assumptions.

\subsection{Viewing effects on cavity morphology}\label{Sec:Effect of viewing on cavity topology}
\begin{figure*}
\centering
	\includegraphics[scale=0.45]{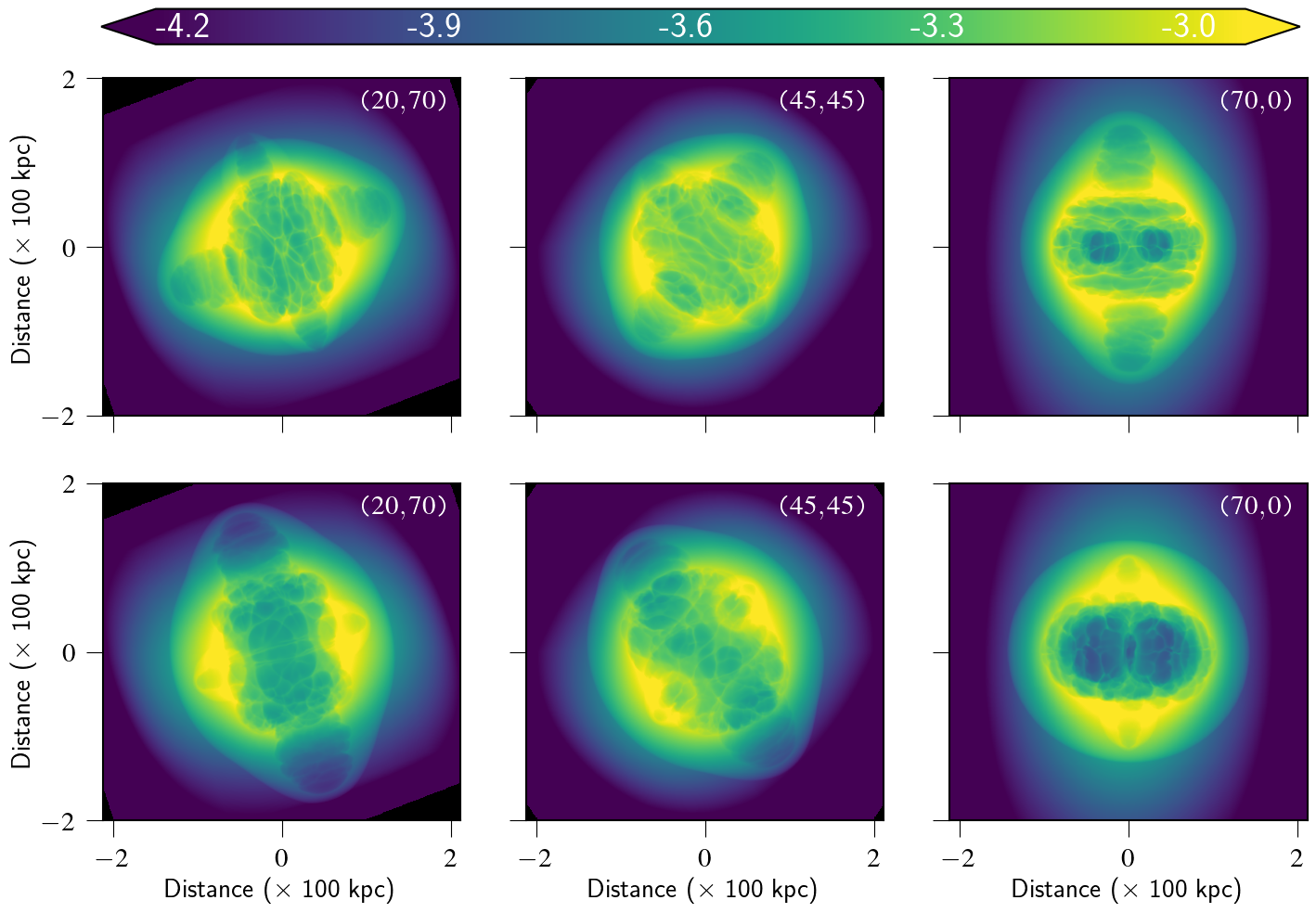}
    \caption{Numerous cavity morphologies (at 114 Myr), obtained for the case of `qr{\small \texttt{90}}\rule{2mm}{.5pt}{\small YX}' (\textit{top panel}) and `qr{\small \texttt{90}}\rule{2mm}{.5pt}{\small XY}' (\textit{bottom panel}) for  three different line of sight angles $(\theta,\phi) \equiv (20^{\circ},70^{\circ}),\ (45^{\circ},45^{\circ}),\ (70^{\circ},0^{\circ})$ (denoted in the respective figures). The colorbar complements the intensity distribution plotted in ${\rm log[erg\ s^{-1} cm^{-2} sr^{-1}]}$. Such complex cavity geometry may look distinctive but is not uncommon. It is interesting to note the appearance of four (mostly elongated) X-ray cavities in the emission maps associated with the reorientation cases irrespective of the line of sight visualization angles.}
    \label{fig:Cavity_LOS}
\end{figure*}

It is important to consider how the X-ray cavities in Fig.~\ref{fig:Cavity_evolution} appear depending on the selection of the
viewing angle. 
In order to further investigate the projection effects, we have chosen three different combination of viewing angles, 
$(\theta,\phi)\equiv(20^{\circ},70^{\circ}),(45^{\circ},45^{\circ}),(70^{\circ},0^{\circ})$ \citep{Giri2022A},
for the case `qr{\small \texttt{90}}\rule{2mm}{.5pt}{\small YX}' and `qr{\small \texttt{90}}\rule{2mm}{.5pt}{\small XY}'. 
In Fig.~\ref{fig:Cavity_LOS}, we depict the corresponding cavity morphologies (at 114 Myr) in the top and bottom panel, respectively.

For the case `qr{\small \texttt{90}}\rule{2mm}{.5pt}{\small YX}', the X-ray cavity structure obtained from the line of sight viewing angles
$(20^{\circ},70^{\circ})$ and $(45^{\circ},45^{\circ})$ show resemblance to the cavity structure observed in Fig.~\ref{fig:Cavity_evolution}. 
These maps clearly show an extended cavity in the center of the cluster surrounded by an illuminating rim of enhanced
emission, as well as four noticeable elongated cavities further away.
The projection effect, however, diminishes the prominence of the outside elongated cavities in the later scenario, i.e., for the $(45^{\circ},45^{\circ})$ case. 
A nearly comparable impact for a perpendicularly reorienting jet on the ambient medium is observed in NGC 193 \citep{Bogdan2014}, implying that
such structures associated with XRGs could indeed be expected.

In this regard, the morphology produced for the $(70^{\circ},0^{\circ})$ viewing angle is particularly intriguing, where the structure is viewed 
nearly along the active lobe direction. 
As a result of projection, the cavity connected to the active lobe merges with the X-ray dip region belonging to the thick cocoon
in the middle, forming two nearly circular and prominent X-ray holes. 
The cavity pair associated with the wing structure is most prominent in this case, as it is not affected by projection effects. 
This scenario with $(70^{\circ},0^{\circ})$ projection direction shows how active jets pierce through the surrounding medium, making a hollow channel
that the (back-flowing) material can easily move through.
This cavity structure also exhibits how material from the active jet flows back into the older channel along the wing, which was made by the earlier jet, keeping the wing structure active and growing at a supersonic speed.

For all these different viewing angles, the outer shock surface, generated by the bow shock from the older jet,
is visible \citep[similar to][]{Cielo2018}.  

The X-ray maps for case `qr{\small \texttt{90}}\rule{2mm}{.5pt}{\small XY}' for the same projection direction are illustrated in the lower panel of Fig.~\ref{fig:Cavity_LOS}. 
In this case also, for the $(20^{\circ},70^{\circ})$ and $(45^{\circ},45^{\circ})$ scenarios, the appearance of similar cavity morphology is observed (similar to Fig.~\ref{fig:Cavity_quantification}; 4th row). 
The occurrence of such structures is not surprising, as it is within the expectations  
of anticipated outcomes from reorienting jets \citep[e.g.,][]{Hodges-Kluck2010B,Chon2012,Cielo2018,Lalakos2022}. 
The morphology of the cavity structure connected with the later scenario although causes some uncertainty 
in predicting the origin (Back-flow or Jet-reorientation).
This uncertainty is due to the fact that the cavities, especially the pair that is associated with the active jet, merge into the central X-ray depression region.

The example with $(70^{\circ},0^{\circ})$ viewing angle is particularly interesting.
Here, we look at the jet-cocoon induced cavity structure along the wing. 
From this perspective, a nearly circular and more noticeable cavity pair is detected, along with a perpendicular cavity pair accompanying the active jet head. 
A recently discovered X-ray cavity system in the galaxy cluster RBS 797 \citep{Ubertosi2021} further demonstrates the existence of analogous example 
in the deep exposure X-ray map, validating the Jet-reorientation hypothesis from their multi-band observations.
This geometrically complex cavity distribution is surrounded by a shocked surface further out, as also seen in \citet{Ubertosi2022}, resembling our 
acquired X-ray map, which was formed as a result of the older jet's expansion and consequently the wing expansion.

\subsection{Estimating the Dynamical Age}\label{Sec:Temporal analysis}
The temporal evolution of shocks generated by active lobes and wings is a critical factor in comprehending and comparing the dynamic evolution of these structures.
We evaluate the shock ages following \citet{Randall2011,Ubertosi2022}, 
\begin{equation}
    {\rm t_{age}} = \frac{R}{\mathcal{M_S}c_\mathcal{S}}
\end{equation}
by tracking the associated bow shock locations. 
Here, $R$ is the projected mid-aperture distance of the forward shock associated with the lobes to the centre of the cluster, $c_\mathcal{S}$ is the speed of sound in the intra-cluster medium,
and $\mathcal{M_S}$ is the Mach number of the forward shocks (noted in Section~\ref{Sec:Dynamical evolution: Development of the wings}). This age represents the time required for the forward shocks to reach their current locations from the core, which is a proxy of the dynamic age of the lobes.

We calculate the average age of shocks associated with the active lobe $<{\rm t_{age}^{l}}>$ and the wing $<{\rm t_{age}^{w}}>$. 
We consider two cases, `qr{\small \texttt{90}}\rule{2mm}{.5pt}{\small YX}' and `qr{\small \texttt{90}}\rule{2mm}{.5pt}{\small XY}' 
(see Fig.~\ref{fig:Cavity_quantification},~\ref{fig:Cavity_LOS} for the projected bow shock locations), where the jets performed re-orientation instantaneously. The age of shocks due to active lobes and wings from these cases are presented in Table~\ref{Tab:Age}. 
It can be noted that the estimated age of the shocks exceeds the actual dynamical age, regardless of projection effects, except in cases of extreme viewing angles.
Such an overestimate can be attributed to the higher expansion speed of shock during the early course of their development.
Further, this technique exhibits the ability to discern older shocks, barring instances of extreme viewing angles.

The estimated shock ages are listed in  Table~\ref{Tab:Age} along with their variation with  viewing orientations. 
In the present work, the actual (dynamical) age of the older lobe, i.e., the wing is 114 Myr, and of the active younger lobe is 36 Myr.
For extreme viewing angles, such as $(70^{\circ}, 0^{\circ})$, where we observe the structure close to one pair of lobes, the forward shock associating 
the lobes blends in such a way that the age estimation is affected. 

Although this method yields a comparable result in terms of shock age when compared to the actual dynamical age associated with the wing (114 Myr), it
deviates by nearly 1.7 to 2.7 times when estimating the age of the active lobe. 
This deviation also appears due to the jet's rapid propagation through the existing cocoon after flipping to the newer direction, as the heavier ambient material 
has already been pushed aside by the older over-pressured cocoon (see Section~\ref{Sec:Dynamical evolution: Development of the wings}).

\begin{table}
\caption{We show here the age of forward shocks evaluated for the cases `qr{\small \texttt{90}}\rule{2mm}{.5pt}{\small YX}' and `qr{\small \texttt{90}}\rule{2mm}{.5pt}{\small XY}', as appeared for different viewing directions (2nd column). The average shock ages associated with the active lobe and the wing are shown in 3rd and 4th columns, respectively. }
\label{Tab:Age}
\centering
\begin{tabular}{ l|c|c|c }
 \hline
 \centering {\centering Simulation label}&Viewing angle& $<{\rm t_{age}^{l}}>$ & $<{\rm t_{age}^{w}}>$ \\
  &$(\theta,\phi)$ in degree& (Myr) & (Myr) \\
 \hline
 \multirow{4}{0.6cm}{\rotatebox[origin=c]{0}{qr{\small \texttt{90}}\rule{2mm}{.5pt}{\small YX}}}&$(0, 0)$& $75.6$ & $127.2$ \\ 
 &$(20, 70)$& $67.0$& $135.1$ \\ 
 &$(45, 45)$ & $60.3$  & $127.0$ \\
 &$(70, 0)$ & $26.7$  & $141.1$ \\
 \hline
 \multirow{4}{0.6cm}{\rotatebox[origin=c]{0}{qr{\small \texttt{90}}\rule{2mm}{.5pt}{\small XY}}}&$(0, 0)$& $98.9$ & $130.3$  \\ 
 &$(20, 70)$& $92.3$& $129.3$ \\ 
 &$(45, 45)$ & $87.1$  & $118.2$ \\
 &$(70, 0)$ & $97.7$  & $73.5$ \\
 \hline
\end{tabular}
\newline
\newline
\raggedright Note: The older and younger forward shocks, associated with the wings and active lobes, possess ages 114 Myr and 36 Myr, respectively.
\end{table}

\section{Summary}\label{Sec:Summary}
The objective of this study is to better understand the origin and long-term evolution of X-shaped radio galaxies (XRGs) in a galaxy cluster environment. 
In this regard, we have performed relativistic MHD simulations considering the jet Back-flow and Jet-reorientation scenarios leveraging the adaptively refined mesh with the {\sc PLUTO} code. 
For each of the simulation runs conducted, we have analyzed the jetted structure as well as its effect on the surrounding cluster medium.
The jetted structure was investigated by examining the dynamical structures that emerge over time, whereas the ambient medium was investigated by deriving
thermal Bremsstrahlung (X-ray) emission maps (0.5-5 kev). 

The results obtained from our study are summarized below:
\begin{enumerate}[label={{\roman*.}}, leftmargin=1.0em,itemsep=2pt, parsep=2pt, topsep=0pt]
    \item This study reveals that each of the XRG formation models we employed has the ability to generate this particular winged radio galaxy, which also has morphological 
    resemblances to various observed XRGs. 
    This finding provides compelling evidence that there may not exist a universal mechanism for the formation of XRGs.
    \item The Jet-reorientation model appears to provide a natural explanation for certain noteworthy yet disputed features of XRGs, as opposed to the intricate involvement 
    of multiple parameters from the Back-flow model. The Jet-reorientation model can explain the XRGs whose wings are aligned along the major axis of the ambient medium, explains the formation of well-collimated wing and active lobe structures, and can naturally produce wings that are noticeably longer than the active lobes. 
    Our results provide support for the idea that a combination of the aforementioned three features, as detected in a small number of XRGs, plausibly favors a scenario of a jet reorientation for their origin. The wing structures, regardless of the models applied, are observed to propagate supersonically throughout the course of their evolution.
    \item Based on the comparable XRG shapes resulting from the Jet-reorientation scenarios involving rapid and gradual flip processes, it can be inferred that a 5 Myr jet reorientation
    period is still fast enough to provide the final structure any distinguishable features. 
    This parameter of the jet reorientation time, as sensed in our study, has become increasingly significant in elucidating various winged radio galaxies, including S- or Z-shaped sources. 
    \item We observed that XRGs resulting from the Jet-reorientation scenario exhibit greater potential for disrupting the ambient medium. 
    This result is evidenced by the generation of multiple, isotropic shock waves, resulting in an extensive dispersion of energy into the surrounding environment, more as compared to the Back-flow scenario. 
    \item This study's findings further suggest that XRGs generated in any of the models investigated, develop a turbulent cocoon structure.
    The momentum components of the plasma material within the jet-inflated cocoon vary across the different models that we have simulated. 
    We suspect that these observed differences in momentum components may result in different levels of turbulence, which may then influence the non-thermal emission markers of said structures.   
    \item The thermal X-ray map of the ambient cluster medium provides evidence of the close relationship between the jet propagation and the ambient medium, 
    as evidenced by the existence of cavities and bright rim signatures.
    From our simulations we obtain four cavities in the X-ray maps, irrespective of the models, associated with wing and active lobe. 
    \item Regarding model differentiation, XRGs emerging from Jet-reorientation scenarios exhibit elongated cavity structures that align with the lobes' propagation direction.
    Additionally, the cavities linked to the wings  show a larger contrast with the surroundings than those associated with the active lobes. 
    Conversely, the XRGs originating from the Back-flow model display a higher cavity contrast in the active lobe than in the wing. 
    Additionally, the cavities linked to the wing have been observed to possess an elongated shape that is perpendicular to the direction of wing growth. 
    We note that all the cavities detected in this study are {`}clear{'} cavities with a fractional difference in X-ray emission $> 35\%$ compared to
    the surrounding region. 
    \item The extent of the cavity depends on the observed line of sight. The cavities that are linked to wings or active lobes, tend to blend or merge with other pre-existing cavities for situations of extreme viewing angles. Nevertheless, in all of these circumstances, we were able to unambiguously identify the presence of four distinct regions of decreased X-ray emission. 
    The acquired cavity distributions do not appear to be uncommon from an observational point of view. 

    \item The effect of the projection on the observed XRGs morphology is further anticipated to influence the estimation of other parameters, such as determining the shock ages when employing observational approaches. 
     Such methodologies also appear to neglect various inherent phenomena that affect the propagation of jets. 
     Irrespective of these effects, the observational method we used in this study shows the capability of identifying the older and younger shocks, hence the lobes, 
     in most of the scenarios.

\end{enumerate}

\noindent Overall, by utilizing numerical simulations of jet interaction with the ambient medium, we have demonstrated that radio jetted structures with peculiar morphologies
have a substantial impact on the ambient medium.
We show that the combination of both thermal X-ray maps of the ambient medium and the dynamical structures mocking radio cocoon can indeed 
provide a comprehensive understanding of the underlying formation mechanisms of such unusual jetted sources. \\

We thank an anonymous referee for offering helpful suggestions to enhance the presentation of the paper content. 
GG is supported by the Prime Minister's Research Fellowship, and is grateful for the financial aid granted in carrying out this project.  
BV is leading the Max Planck Partner Group at Indian Institute of Technology Indore (IIT Indore), India. 
CF acknowledges travel funds by the Max Planck Partner Group at IIT Indore.
The computational modeling presented in this study was performed using the Max-Planck-Gesellschaft Super-computing Cobra cluster
(\url{https://www.mpcdf.mpg.de/services/supercomputing/cobra}) and using resources available at the IIT Indore.

%% Also note that the akcnowlodgment environment does not support long amounts of text. If you have a lot of people and institutions to acknowledge, do not use this command. Instead, create a new \section{Acknowledgments}.


\appendix
\section{Large-angle reorientation of the jet axis} \label{Sec:Large-angle reorientation of Jet axis}
As a most plausible scenario for reorienting the jet ejection axis to a significant angle is by similarly large reorientation of the black hole's spin axis.
A number of analytical studies have explicitly investigated the potential conditions under which such events may occur. 

The processes attributed to cause a substantial change in the black hole spin axis include 
(i) the coalescence of two black holes, or 
(ii) a rapid mass inflow into the central black hole, misaligned with the existing accretion disk. 
In this context, we briefly discuss two analytical studies below, which are relevant for the phenomenon of X-shaped Radio Galaxies (XRGs).

\citet{Merritt2002} investigated the merger of two supermassive black holes applying a mass ratio of $M_1/M_2 = 1/4$, exploring the change in spin magnitude $\delta S$ during the coalescence, 
which can be approximated as $GM_1^2/2c$. 
Here, $\delta S = S - S_1$, where $S$ denotes the resultant spin after the coalescence of the black holes,
while $S_1$ denotes the spin of the heavier black hole involved in the merger.
Introducing the parameter $\lambda = \delta S/S_1$, the study reveals that the average cosine of the realignment angle $\zeta$ 
can be expressed as $\langle \cos\zeta \rangle \simeq (2/3\lambda)$ for $\lambda > 1$. 

Notably, if the larger black hole is rotating initially at a relatively slow rate (compared to its maximum value $GM_1^2/c$), for example $\sim 0.1\,G M_1^2/c$ (thus $\lambda > 1 $),
the derived realignment angle is approximately $\simeq 82^{\circ}$. 
An intriguing result from the recent study conducted by \citet{Garofalo2020} suggested the existence of low spinning black holes in X-shaped radio galaxies, 
by attributing XRGs as transition objects between low-spinning retrograde and prograde black holes.

In the context of inhomogeneous mass accretion onto supermassive black holes, the study by \citet{Babul2013} proposes a relation connecting the maximum tilt angle $\zeta_{\rm max}$ to
the ratio between gas inflow $\Delta M_{\rm gas}$ and the black hole mass $M_{\bullet}$, 
 \begin{equation}
 \zeta_{\rm max} \approx \tan^{-1} \left( 73.5 \frac{\Delta M_{\rm gas}}{M_{\bullet}} \right)
 \end{equation} 
When considering a relatively modest gas inflow, such as $\Delta M_{\rm gas} \approx 3 \times 10^7 M_{\odot}$,
the authors estimate a resulting tilt angle of approximately $\sim 65^{\circ}$ for $M_{\bullet} = 10^9 M_{\odot}$. 
It is important to note that this magnitude of the tilt angle is specifically applicable to slowly rotating black holes.

Further studies, for example in \citet{Gergely2009,Lalakos2022} also lend support to these ideas, in particular concerning X-shaped radio galaxies. 
The former study associated the formation of wing structures in XRGs with binary SMBH coalescence, considering mass ratios ranging from 1/3 to 1/30. 
The later study proposed that the jet ejection axis flips to a large angle due to inhomogeneous mass accretion, leading to erratic wobbling.

Thus, the aforementioned studies provide support for the feasibility of large-angle jet reorientation in XRGs. 
However, it is essential to conduct further investigations as a number of assumptions and parameters involved in these studies are not yet fully constrained.
For instance, while \citet{Joshi2019} provided some initial estimates of black hole masses in a statistical sample of X-shaped radio galaxies (XRGs), 
the understanding of the gas inflow rates to the central black hole responsible for a jet reorientation remains incomplete. 

Additionally, a recent study by \citet{Giri2023} revealed that most, if not all, elliptical galaxies in the nearby universe undergo mergers, suggesting that this 
phenomenon plays a crucial role in the hierarchical evolution of galaxies. 
Therefore, it is evident that not every galaxy merger event may efficiently reorient the spin axis of a supermassive black hole by a significant angle.
Such observations are further corroborated by the scarcity of X-like shapes in weaker, edge-dimmed FR-I radio sources \citep{Dennett-Thorpe2002,Gopal-Krishna2012}. 
All these pivotal aspects definitely warrant further exploration and verification.


\bibliography{sample}{}
\bibliographystyle{aasjournal}



\end{document}